\documentclass[a4paper,12pt]{article}
\usepackage{psfrag}
\usepackage{graphicx}
\usepackage{graphics}
\usepackage{bm}
\usepackage{color}
\usepackage{amssymb}
\usepackage{amsmath}
\usepackage{enumitem}
\input{epsf.sty}
\usepackage{graphicx}

\usepackage{vmargin}            % redŽfinir les marges
\setmarginsrb{2cm}{2cm}{2cm}{2cm}{0cm}{0cm}{0cm}{1cm}

\newcommand\ba{{\mathbf{a}}}
\newcommand\bb{{\mathbf{b}}}

\newcommand\bx{{\mathbf{x}}}

\newcommand\bX{{\mathbf{X}}}

\newcommand\be{\begin{equation}}
\newcommand\ee{\end{equation}}
\newcommand\beq{\begin{equation}}
\newcommand\eeq{\end{equation}}
\newcommand\bea{\begin{eqnarray}}
\newcommand\eea{\end{eqnarray}}

\newcommand\si{\sigma}

\def\d{{\rm d}}
\def\n{\mathbf{n}}
\def\b{\mathbf{b}}
\def\a{\mathbf{a}}
\def\B{\mathbf{B}}
\newcommand{\gsim}{\raise.3ex\hbox{$>$\kern-.75em\lower1ex\hbox{$\sim$}}}
\newcommand{\lsim}{\raise.3ex\hbox{$<$\kern-.75em\lower1ex\hbox{$\sim$}}}

\def\vf{}

\title{Gravitational wave bursts from cosmic superstrings with Y-junctions}
\author{P.~Bin\'etruy\footnote{binetruy@apc.univ-paris7.fr} , A.~Boh\'e\footnote{bohe@apc.univ-paris7.fr} , T.~Hertog\footnote{hertog@apc.univ-paris7.fr} 
  and D.A.~Steer\footnote{steer@apc.univ-paris7.fr} \\
{\small {}}\\
{\small {\it  APC\ \footnote{Universit\'e 
Paris-Diderot, CNRS/IN2P3,  CEA/IRFU and Observatoire de Paris} ,10 rue Alice Domon et L\'eonie Duquet,
 75205 Paris Cedex 13, France}}\\
}

\begin{document}
\date{\today}
\maketitle

\begin{abstract}

%We revisit the gravitational wave bursts emitted by kinks and cusps on cosmic 
%superstring loops. Such loops generically contain strings of different tensions
%which meet at Y-junctions. These resulting loops evolve non-periodically in 
%time, but still contain cusps and kinks which now interact with the junctions 
%in important ways.  We set up the formalism to study the gravitational wave 
%bursts from such non-periodic  line-like sources. We find that, besides the 
%standard cusps of kinks, which remain mostly unchanged from the standard case, 
%we find new types of contributions: (i) strings expanding at the speed of 
%light at a junction, (ii) kinks passing through a junction. \\
%OR\\

Cosmic superstring loops generically contain strings of different tensions that meet at Y-junctions. These loops evolve non-periodically in time, and have cusps and kinks that interact with the junctions. We study the effect of junctions on the gravitational wave signal emanating from cosmic string cusps and kinks.
We find that earlier results on the strength of individual bursts from cusps and kinks on strings without junctions remain largely unchanged, but junctions give rise to additional contributions to the gravitational wave signal coming from strings expanding at the speed of light at a junction and kinks passing through a junction. 

\end{abstract}

%\pacs{11.27.+d,98.80.Cq}

%\maketitle

\section{Introduction}

In string theory models of brane inflation, fundamental (F) and D-strings are produced at the end of inflation when the branes collide and annihilate \cite{Polchinski:2004hb,Kibble:2004hq,Davis:2005dd,Copeland:2003bj}. In certain scenarios the resulting superstring networks are stable and expand to cosmic size, with predicted string tensions in the range $10^{-11} \leq G\mu \leq 10^{-6}$. This raises the possibility that cosmic superstrings could provide an observational signature of string theory. Indeed it has been argued that the gravitational wave (GW) signals from cusps of oscillating loops on cosmic strings should be detectable by LIGO and LISA for string tensions as low as $10^{-13}$  \cite{DVletter,DV,DV1}.

However the GW predictions of \cite{DV} may not be directly applicable to cosmic superstring networks, because F and D-strings form an interconnected network in which they join and separate at Y-junctions. Each junction joins an F-string, a D-string and 
%a (1,1) 
their bound state.
% of F and D. 
This means that closed loops containing junctions in a cosmic superstring network do not evolve periodically in time.
Furthermore kinks interact with junctions, which leads to novel contributions to the GW signal as well as a more complicated network evolution. For these reasons we re-examine here the gravitational wave signal emanating from cusps and kinks on cosmic string loops taking in account
the presence of junctions. We note that in a similar spirit, Damour and Vilenkin %{\bf DS: I still feel we should cite Siemens et al somewhere here}%
have calculated the effect on GW signatures of cosmic strings of a reduced reconnection probability of intersecting strings and of a reduced typical length of newly formed loops. It was found that earlier results obtained for field theory cosmic strings remain largely valid for a rather wide range of network parameters.

We first set up the formalism for calculating GW bursts emitted by cusps and kinks on non-periodic cosmic string loops. 
The non-periodic evolution of loops with junctions renders the GW calculation somewhat more involved because one can no longer factorize the Fourier transform of the GW amplitude. We show one can nevertheless integrate the stress energy over the string worldsheet and obtain an analytic expression for the high frequency behavior of the various contributions to the GW bursts. We find that earlier results on the strength of individual bursts from cusps and kinks on strings without junctions remain largely unchanged, but junctions give rise to additional contributions coming from strings expanding at the speed of light at a junction and kinks passing through a junction. We analyze the latter contributions, which provide a possible observational discriminant between ordinary cosmic strings and cosmic superstrings.

We concentrate here on the calculation of individual gravitational bursts. This has the advantage of being a fully tractable exercise. By contrast, the observable signal of such bursts at the present time depends also on the details of the loop evolution as well as on the cosmological evolution of the network. In particular junctions tend to enhance the number of kinks on loops. One might expect this amplifies the GW signal from cosmic superstring networks compared to the signal from cosmic string networks without junctions. 
A more detailed analysis of this effect will appear elsewhere \cite{BBHS2}.

\section{GW emission from cosmic superstrings with junctions}
\label{CSwJ}

In this section we consider the GW emission from a cosmic superstring loop, 
and generalize the results of  Damour and Vilenkin \cite{DVletter,DV,DV1}
on the emission of gravitational bursts from standard loops without junctions.

\begin{figure}[h] %  figure placement: here, top, bottom, or page
   \centering
   \includegraphics[scale=0.3]{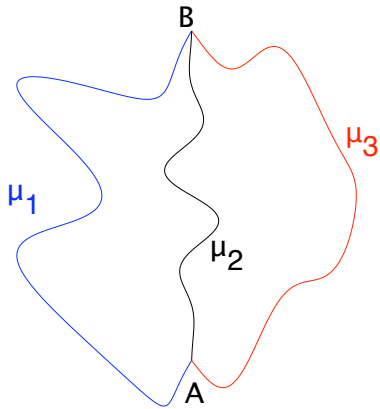} 
   \caption{{\bf Loop formed by 3 strings and 2 junctions}}
   \label{fig:loop}
\end{figure}

We consider closed loops of cosmic superstrings containing two Y-junctions, as shown in figure  \ref{fig:loop}.  (Generalization to loops with 4 or more junctions is straightforward in principle). 
%%%DS%%%
The loop consists 
of three local Nambu-Goto cosmic strings of tensions $\mu_q$ $(q=1,2,3)$, 
which meet at two Y-junctions where the junctions are labeled by $A$ and $B$. 
The action describing the system has been set up and analysed in \cite{CKS}, 
and is given by
 \be
 S=-\sum_{q=1,2,3}\mu_q \int \d t \int_{s^A_q(t)}^{s^{B}_q(t)} \d\si_q\, 
\sqrt{{\bx'}_q^2 \, (1-\dot \bx_{q}^{2}) } 
+\sum_{J=(A,B)}\sum_q\int \d t \,
 {\mathbf f}^J_{q}  \cdot [\bx_{q}( t,s^J_q (t) )-{\bX}_J(t)].
 \label{action1} 
 \ee
The first term is the Nambu-Goto action for each of the three strings in the 
loop, where we have assumed a flat spacetime geometry 
%why flat space?
with signature $(-+++)$ 
and used the standard conformal-temporal gauge to parametrize each string's 
worldsheet. Thus each string is described by its spatial coordinates $\bx _{q}
(\sigma_q,t)$ where $t$ coincides with background time.   (As discussed in 
\cite{Copeland:2007nv}, cosmic superstrings should be described by the 
Dirac-Born-Infeld action, but the resulting equations of motion reduce to those derived from (\ref{action1}).)
Since each string is bounded by the two junctions, the $\sigma$ parameter runs 
between two time-dependent bounds
\be
\label{ }
\sigma_q \in [s _{A,q}(t),s _{B,q}(t)].
\ee
The conformal gauge constraints can be written (with $'$ and $\cdot$ standing 
for derivatives with respect to  $ \sigma$ and $t$) as
\bea
\bx'  _{q}\cdot \dot{\bx } _{q}& = & 0 \\
\bx_{q}'^{2} + \dot{\bx } _{q} ^{2} & = & 1
\label{gauge}
\eea
and as usual, away from the junctions, the wave-like equation of motion  
$\ddot\bx_q-\bx_q''=\bf{0}$ yields
\be
\label{ }
\bx _{q}(\sigma,t)=\frac{1}{2}(\ba_{q}(\sigma+t)+\bb_{q}(\sigma-t)),
\ee
with $\ba_{q}'^{2}=\bb_{q}'^{2}=1$ in order to satisfy the gauge constraints.
This resembles the solution for standard closed loops. 
However, one must also take in account the second term of (\ref{action1}) which imposes, via the Lagrange multipliers 
$ {\mathbf f}^J_{q}$, that the 3 strings meet at the 
positions of the junctions $\bX_A(t)$ and $\bX_B(t)$.  Causality requires
$|\dot{\bX}_J(t)|\leq 1$, which together with (\ref{gauge}) yields \cite{CKS}
\be
|\dot{s}_{A,q}|\leq1 \, ,\qquad |\dot{s}_{B,q}| \leq 1.
\label{causality}
\ee
The time evolution of $s_{A,q}(t)$ and $s_{B,q}(t)$ is given by the equations of motion at the junction, yielding
%In the case considered in (\ref{action1}), one obtains 
\cite{CKS} 
\be
\label{s1}
{\mu_1 (1-\dot s_{B,1}) \over \sum_q \mu_q} = {M_1(1-c_{B,1}) \over \sum_q M_q(1-c_{B,q})}
\ ,
\end{equation}
where
\be
\label{c1M1}
c_{B,1} \equiv \bb'_2(s_2-t)\cdot \bb'_3(s_3-t) \quad , \quad 
M_1 \equiv \mu_1^2 - (\mu_2-\mu_3)^2 \geq 0\ ,
\ee
and similarly by circular permutation.
As a result of the presence of the junctions, loops now  evolve non-periodically in time.

{\vf Like field theory cosmic strings, superstrings can contain cusps --- points moving at the speed of light 
$|\dot{\bx}_q| = 1$ (and hence $\bx_q'=0$) with $\bx_q'' \neq 0$ 
\cite{ViSh,HiKi} --- and kinks, which correspond to a discontinuity in $\bx'_q$. 
Here we aim to study the gravitational wave emission by such localized sources. 
We first calculate the GW signal in the local wave zone of the source, at distances from the source that are larger than the wavelength of interest but smaller than the Hubble radius. In this regime we can take the spacetime to be asymptotically flat:
$g _{\mu\nu}=\eta _{\mu\nu}+h  _{\mu\nu}$, where $h_{\mu \nu} \ll 1$ is the metric perturbation generated by the source. The subsequent propagation of the gravity waves on cosmological scales in a Friedmann-Lemaitre spacetime is discussed at the end}.

%\\
%\label{ }
%h _{ij}&=&\tilde{g} ^{(1) \scriptscriptstyle(TT)} _{ij} \, .
%\eea
%

In a suitable gauge the GW are described by the transverse traceless $^{\scriptscriptstyle(TT)}$ part of the linear perturbation $h _{ij}^{\scriptscriptstyle(TT)}$ of the spatial metric. This satisfies the linearized Einstein equations, 
\be
\label{ }
\square h _{ij}^{\scriptscriptstyle(TT)}  = -16 \pi G T _{ij} ^{\scriptscriptstyle(TT)} \, .
\ee
where $T _{ij} ^{\scriptscriptstyle(TT)}$ is the $^{\scriptscriptstyle(TT)}$ part of the stress-energy tensor of the source.
In the local wave zone, for closed loops of characteristic size $L$ localized around the origin, 
the solution is given by \cite{DV} 
\be
\label{Tijtohij}
h _{ij}^{\scriptscriptstyle(TT)}
(\bx,\omega)=\frac{4G}{r}e ^{i\omega r}T _{ij} ^{\scriptscriptstyle(TT)}
(\omega \mathbf{n},\omega)
\ee
where $\mathbf{n}=\mathbf{x}/\|\mathbf{x}\|$ and $r=\|\mathbf{x}\|$. Note that 
$h _{ij}^{\scriptscriptstyle(TT)}$ is actually the time Fourier transform of $h_{ij}^{\scriptscriptstyle(TT)}$ and $T _{ij} 
^{\scriptscriptstyle(TT)}$ the space-time Fourier transform of $T _{ij} 
^{\scriptscriptstyle(TT)}$: we will use these abusive notations throughout the 
text. 
%Let $L$ be the characteristic size of the loop. 
The wave-zone conditions are \cite{Weinberg,Maggiore:1900zz}
\be
\label{wave-zone}
r  \gg  L \nonumber \ , \qquad
r  \gg  L ^{2} \omega \ , \qquad
r  \gg  1/\omega  \ .
\ee
%We thus see that the only ingredient needed to study the gravity waves 
{\vf Hence to obtain the GW emission in this regime
it suffices to evaluate $T _{ij}^{\scriptscriptstyle(TT)}$ at points $(\mathbf{k},\omega)$
in the Fourier domain that satisfy the dispersion relation $\mathbf{k}=\omega
\mathbf{n}$, where $\mathbf{n}$ is a unit vector pointing from the source towards the observer.}

%(far from the source and in the range of frequencies allowed by the previous conditions) is the $^{\scriptscriptstyle(TT)}$ part of $T _{ij}$. To be more precise, we 

Let us
%%%DS%%% momentarily  TH:??
put aside the $^{\scriptscriptstyle(TT)}$ projection 
%{\vf TH:why? don't we need to project eventually? are the novel contributions we find TT?} 
From now on we concentrate on the calculation of $T _{ij}$ and merely indicate where the $^{\scriptscriptstyle(TT)}$ projection projects out a leading contribution. From the action (\ref{action1}) one gets \cite{CKS}
%. For $a$ and $b$ spatial indices,
%
\be
\label{ }
%\begin{split}
T ^{ij}(\mathbf{x_0},t_0)=\sum _{{\rm string}\ q} \mu_q\int \d t 
\int _{s _{A,q}(t)} ^{s _{B,q}(t)}
%& 
\left. (\dot{x}^i_q \dot{x}^j_q-x^{\prime i}_q  x^{\prime j}_q) 
\right|_{(\sigma,t)}
% \\
%&\times
 \delta ^{(3)} (\mathbf{x_0}-\mathbf{x}_q (\sigma,t))\delta(t_0-t) \d\sigma 
\, ,
%\end{split}
\ee 
namely a sum of contributions of all strings. Each term can be evaluated 
independently once the coupled dynamics of the system has been determined. 
From now on, we focus on one of these contributions and drop the $q$ subscript 
indicating the string.
Moving into Fourier space, we have
\bea
T _{ij}(\mathbf{k},\omega) & = &  \mu \int \d t \int _{s _{A}(t)} 
^{s _{B}(t)}\left. (\dot{x}_i \dot{x}_j-x^{\prime}_i  x^{\prime}_j) 
\right|_{(\sigma,t)} e ^{i(\omega t - \mathbf{k}\cdot\mathbf{x}(\sigma,t))} 
\d\sigma
\\
 & = & \frac{\mu}{2} \int \d t \int _{s _{A}(t)} ^{s _{B}(t)} a^{\prime} 
_{(i}  (\sigma+t) \; b^{\prime} _{j)}(\sigma-t) e ^{i(\omega t - 
\mathbf{k}\cdot\mathbf{x}(\sigma,t))} \d\sigma
 \label{stressE}
\eea

\begin{figure}[h] %  figure placement: here, top, bottom, or page
   \centering
   \includegraphics[scale=0.4]{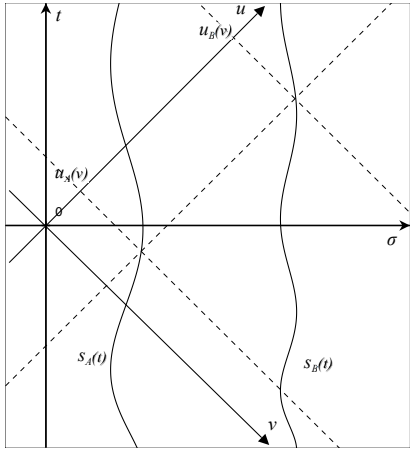} 
   \caption{{\bf Worldsheet of one of the strings}}
   \label{fig:worldsheet}
\end{figure}

For the periodic loops with no junctions considered in \cite{DV}, changing  
integration variables to
\be
\label{ }
u= \sigma+t \ , \qquad  v= \sigma-t \ .
\ee
and using the periodicity leads to a {\it factorized} expression of the form
\be
\label{fpcase}
T _{ij}\propto \Big( \int   a^{\prime} _{(i}(u) e ^{\frac{i}{2}(\omega u -  
\mathbf{k}\cdot\mathbf{a}(u))} \d u \Big) \Big( \int  b^{\prime} _{j)}(v) 
e ^{-\frac{i}{2}(\omega v +  \mathbf{k}\cdot\mathbf{b}(v))}  \d v \Big).
\ee
With junctions, the absence of periodicity prevents one from writing the stress-energy in a similar
convenient form. Nevertheless, the same change of 
variables still proves useful to study $T _{ij}$ (see figure \ref{fig:worldsheet}). 
Indeed, after changing to $(u,v)$ in (\ref{stressE}) and evaluating at $\mathbf{k}=\omega\mathbf{n}$
we obtain
\be
\label{nonfstressE}
T _{ij}(\omega,\omega \mathbf{n})= \frac{\mu}{4} \int _{-\infty} ^{\infty} 
b^{\prime} _{j}(v) e ^{-\frac{i \omega}{2}(v +  \mathbf{n}\cdot\mathbf{b}(v))} 
 \Big( \int _{u _{A}(v)} ^{u _{B}(v)} a^{\prime} _{i}(u) e ^{\frac{i \omega}{2}
(u -  \mathbf{n}\cdot\mathbf{a}(u))} \d u \Big) \; \d v .
\ee
Notice that the $u$ and $v$ integrals are {\it no longer factorized} due to 
the $v$ dependence of the bounds in the $u$ integral. Indeed, the bounds
$u _{A}(v)$ and $u _{B}(v)$ are defined as follows. Since $|\dot{s}_A|,
|\dot{s}_B| < 1$, for each $v$ there is a unique $t_A$ and a unique $t_B$ that 
satisfy $s_A(t_A)-t_A=v$, $s_B(t_B)-t_B=v$. Then, $u_A(v)=s_A(t_A)+t_A$ and 
$u_B(v)=s_B(t_B)+t_B$.  

We note for further use that, for $J=A$ or $B$,
\be
\label{dudv}
{du_{J} \over dv} = {\dot s_J + 1 \over \dot s_J - 1}  \leq 0\ .
\ee
%%%DS%%%added \leq
This means that for the ``usual" values of $|\dot{s_J}|$ not too close to 1, 
$du_J/dv$ is negative and of order $1$. 
However,
one can also have the limiting values $du_J/dv=-\infty$ when 
$\dot{s}_J=1$ and $du_J/dv=0$ when $\dot{s}_J=-1$. Since such values
correspond to a junction moving at the speed of light, a phenomeneon which will turn 
out to lead to a gravitational burst, we will consider these more carefully below. 
Note, however, that the equations of motion forbid the strings from shrinking at the speed of light  
(either at junction A: $\dot{s}_A=1$ and $du_A/dv=-\infty$, or at 
junction B: $\dot{s}_B=-1$ and $du_B/dv=0$) \cite{CKS1}. Hence one only needs to consider 
the case of a string expanding at the speed of light, again either at junction A
with $\dot{s}_A=-1$ and $du_A/dv=0$, or at junction B with 
$\dot{s}_B=1$ and $du_B/dv=-\infty$). 

We are interested in the production of gravitational bursts, that is in the 
high frequency $\omega$ regime. We can therefore restrict attention to the 
$\omega \to \infty$ limit of the integrals in (\ref{nonfstressE}), where
standard techniques have been developed to evaluate (\ref{nonfstressE}) \cite{DV,DIS}. 
{\vf We note however that an analysis of this kind does not include the low frequency gravitational 
radiation from the slow motion of the string itself, 
and therefore the resulting stochastic background of gravitational waves arising from this.}

\section{High frequency behaviour of $T_{ij}$: General discussion}
\label{sec:cond}

Let us consider integrals of the form
\be
\label{int}
I(\omega)=\int _{a}^{b} f(t)e ^{-i\omega\phi(t)}\d t
\ee 
in the $\omega \rightarrow \infty$ limit.
%that have quickly varying phases in the following sense:
%$$\dot{\phi}\omega\gg \frac{\dot{f}}{f}$$
We will use the following standard result (see, for example, \cite{DIS}):
%%%DS%%% 
\vspace{0.5cm}
\newpage
{\it If
\begin{enumerate}[label=(\it{\roman{*}})]
\item  $\forall k \geqslant 0$, $  f^{\scriptscriptstyle(k)}(a)
=f^{\scriptscriptstyle(k)}(b) \text{ and }  \phi^{\scriptscriptstyle(k)}(a)
=\phi^{\scriptscriptstyle(k)}(b)$ 
\item $f$ and $\phi$ are $C ^{\infty}$ on $[a,b]$ --- that is, $f(t)$ and
$\phi(t)$ are smooth in the 
interval; and
\item $\forall t \in [a,b]$,  $\dot{\phi}(t)\neq 0$ -- that is, there is no 
stationary phase or saddle points,
\end{enumerate}

then $I(\omega) \rightarrow 0$ as $\omega \rightarrow \infty$, faster than 
any power of $1/\omega$. }

\vspace{0.5cm}
\noindent (To see this, change the variable of integration to $\phi$ using
$(iii)$ and integrate by parts. $(i)$ means the boundary term vanishes. 
Repeating this procedure $N$ times shows that $I(\omega)=O(\omega^{-N})$.)
{\vf Thus integrals of the form \eqref{int} are exponentially small for large values of $\omega$ if
the conditions $(i) - (iii)$ hold and the integrand has a rapidly varying phase, i.e. $\omega
\gg |{\dot{f}}/{f\dot{\phi}}|$.

Integrals of this type appear in the expression for the stress energy,
eq. \eqref{fpcase} and \eqref{nonfstressE}. In the following we will therefore be interested in situations where (at least) one of the above conditions does not hold, since these might lead to high frequency contributions to the stress-energy proportional to small powers of $1/\omega$.} But we first summarize the results obtained in \cite{DV} for 
the factorized stress energy tensor (\ref{fpcase}) of cusps and kinks on periodic strings. In this case the above results
can be applied independently to each integral, and since $(ii)$ is ensured by 
the periodicity, the main contributions to $T _{ij}$ appear when

\begin{itemize}
\item  there is a saddle point in each integral ($(iii)$ violated in both 
integrals), leading to a contribution $T _{ij } \propto 1/\omega ^{4/3}$
and corresponding to the physical situation of a cusp emitting around one 
specific direction;

\item there is a discontinuity of $a'_i$ and a saddle point in the integral 
over $v$ (or vice versa), leading to $T _{ij } \propto 1/\omega ^{5/3}$. 
This is the case of a kink emitting in a one-dimensional fan-like set of 
directions throughout its propagation.

%\item a discontinuity of $a'$ and $b'$ giving $T _{ij } \propto 
%\frac{1}{\omega ^{2}}$ and corresponding to a left-moving and a right-moving 
%kink  meeting on the string

\end{itemize}
All other cases (discontinuities in higher derivatives of $a'$ or $b'$ for 
example) give smaller contributions to $T _{ij}$. 

In the case of loops with junctions we consider here, the stress 
energy tensor given in eq.~(\ref{nonfstressE}) cannot be factorized. It can 
only be written in the form 
\be
\label{nonfstressE2}
T _{ij}(\omega,\omega \mathbf{n})= \frac{\mu}{4} \int _{-\infty} ^{\infty} 
b^{\prime} _{j}(v) e ^{-\frac{i \omega}{2}(v +  \mathbf{n}\cdot\mathbf{b}(v))} 
 I_i(v) \; \d v \ ,
\ee
where
\be
I_i(v):= \int _{u _{A}(v)} ^{u _{B}(v)} a^{\prime} _{i}(u) 
e^{-i \omega \phi(u)}  \d u 
%e ^{\frac{i \omega}{2}(u -  \mathbf{n}\cdot\mathbf{a}(u))} \d u \ .
\label{Iv}
\ee
with
\be
\phi(u)=-\frac{1}{2}(u -  \mathbf{n}\cdot\mathbf{a}(u)) \ .
\label{phiudef}
\ee
%%%DS%%%put defn of \phi here already.
The previous results can only formally be applied to the integral in $I_i(v)$: 
indeed, its bound depend on $v$, so that the $v$-integral in 
(\ref{nonfstressE2}) receives contributions that could prevent it from 
being of the same type. However, as we now discuss, in most situations of interest these 
additional contributions have a simple dependence on $v$, allowing us to generalize the results 
of \cite{DV}.

\section{High frequency behaviour of $T_{ij}$: first integration}

Consider first the integral $I_i(v)$ given in (\ref{Iv}).
In order to be in the rapidly varying phase regime, we require $\omega \gg 
\|\mathbf{a}{''}\|$. If the loop is not too wiggly, one can assume that 
$\|\mathbf{a}{''}\|\sim 1/L$ (recall that $\|\mathbf{a}{'}\|=1$), in which 
case the condition becomes $\omega \gg 1/L$ (note that this is an additional constraint on top of 
the wave-zone conditions (\ref{wave-zone})).
We now examine the high frequency contributions to $I_i(v)$ resulting from the violation of one 
of the conditions listed in the previous section\footnote{An analysis of situations where several conditions are violated is beyond the scope of this work.}.

%%%DS%%%changed some $X$'s to $J$'s below.
%\begin{itemize}
%\renewcommand{\labelitemi}{$\bullet$}
\vskip .4cm
$\bullet$ $(i)$ is violated {\it i.e.~}$\mathbf{a}'(u _{A}(v))\neq\mathbf{a}'(u _{B}(v))$. 
\vskip .3cm

This will generically be the case for a string bounded by junctions. Hence this 
is a novel contribution with respect to \cite{DV}. After integrating by parts, 
the leading order contribution scales as $\omega^{-1}$ and given by
%%%DS%%%old equation in text below
\be
\label{Iedge}
%\boxed{
%I _{edge}(v)\approx \frac{2}{i\omega}\Bigg[\frac{a^{\prime} _{i}(u)}{1 -  
%\mathbf{n}\cdot\mathbf{a}^{\prime}(u)} e ^{\frac{i \omega}{2}(u -  
%\mathbf{n}\cdot\mathbf{a}(u))} \Bigg] _{u=u _{A}(v)}^{u=u _{B}(v)}
%}
I _{i}^{boundary}(v)\approx \frac{2}{i\omega}\Bigg[\frac{a^{\prime} _{i}(u)}{1 -  
\mathbf{n}\cdot\mathbf{a}^{\prime}(u)} e ^{-i \omega \phi(u)} \Bigg] _{u=u _{A}(v)}^{u=u _{B}(v)}
\ee
In {\vf most cases of interest one has} $\left|\frac{du _{J}}{dv}(v)\right|\sim1$, so that
$\frac{a^{\prime} _{i}(u_J(v))}{1 -  \mathbf{n}\cdot\mathbf{a}^{\prime}
(u_J(v))}$ is a slowly varying function of $v$ and
$e ^{-i \omega \phi(u_J(v))}$
is a rapidly varying phase term that must be taken in account when 
performing the integral over $v$ in (\ref{nonfstressE2}), though
the latter remains of the same type as \eqref{int}
\footnote{As mentioned above, when the string expands at the speed of 
light \emph{at junction B}, $\frac{du_B}{dv}$ diverges. One might expect
this leads to a discontinuity-like contribution in the second 
integral. However, both the derivative with respect to $v$ of the 
amplitude $\frac{a^{\prime} _{i}(u_B(v))}{1 -  
\mathbf{n}\cdot\mathbf{a}^{\prime}(u_B(v))}$ and of the phase in the 
exponential diverge. The condition $\dot{\phi}\omega\gg \frac{\dot{f}}{f}$ 
still holds since $\frac{du_B}{dv}$ cancels on both sides. Therefore, no 
discontinuity-like contribution is introduced in the second integral}.

\vskip .4cm
$\bullet$ $(ii)$ is violated in the following way:  there exists a 
$u_*\in ]u_A(v),u_B(v)[$ where $a' _i$ is discontinuous.   (We could also 
consider non regularities in higher order derivatives of $a_i$ but they would 
lead to higher order powers in $1/\omega$.) 
\vskip .3cm

To evaluate this term, we first rewrite
$$I_i(v)=\int _{u _{A}(v)} ^{u_*} a^{\prime} _{i}(u) e ^{{i \omega \phi}} du+\int _{u_*} ^{u _{B}(v)} a^{\prime} _{i}(u) 
e ^{i \omega \phi} \d u ,$$ 
and then integrate once by parts as above to find
\be
\label{Idisc}
I _{i}^{disc}(v)=-\frac{2}{i\omega}\Bigg(\frac{a^{\prime} _{i}(u_* ^{+})}{1 -  
\mathbf{n}\cdot\mathbf{a}^{\prime}(u_*^{+})} -  \frac{a^{\prime} _{i}
(u_*^{-})}{1 -  \mathbf{n}\cdot\mathbf{a}^{\prime}(u_*^{-})}\Bigg)
e ^{i \omega \phi(u_*)}
\ee
As above, we find an $\omega^{-1}$ fall-off. The only dependence on $v$ comes
from the bound $u_A(v)<u_*<u_B(v)$. This means that when we will perform the 
integral over $v$ in (\ref{nonfstressE2}) in section V below, {\vf we will need to 
restrict the $v$ domain of integration appropriately.}

%otherwise the integral over $u$ simply appears as a multiplicative constant.

\vskip .4cm
$\bullet$ {\it (iii)} is violated: there exists\footnote{Strictly speaking, $u_s$ could lie outside $[u_A(v),u_B(v)]$, but close to $u_A$ or $u_B$ and still give rise to a non negligible contribution (cf Appendix A).} a $u_s\in [u_A(v),u_B(v)]$ where the 
phase $\phi(u)$ has a vanishing 
derivative $\phi'(u_s)= -(1-\mathbf{n}\cdot\mathbf{a}'(u_s))/2=0$,
%i.e. $\mathbf{n}\cdot\mathbf{a}'(u_s)=1$ 
%%%DS%%%
or in 
other words $\mathbf{n}=\mathbf{a}'(u_s)$. 
%This would give a {\vf subleading} {\vf TH: instead of degraded, which I'm not sure what it means certainly not in combination with sizeable. correct?}, but possibly sizeable, contribution
\vskip .3cm

In this case the leading contribution to $I_i(v)$
%%%DS%%%
%=\int _{u _{A}(v)} ^{u _{B}(v)} a^{\prime} _{i}(u) e ^{-i\omega\phi(u)}du$ 
comes from the vicinity of the saddle point and is obtained by Taylor 
expanding $\phi(u)$ and $a'_i(u)$ around $u_s$. 

To do so, notice that the gauge conditions (on the worldsheet) enforce $ \phi''(u_s)=0$. Indeed, 
$\phi''(u)=\frac{1}{2}\mathbf{n}\cdot\mathbf{a}''(u)$ and since 
$\mathbf{a}'(u)^2=1$, it follows that $\mathbf{a}'\cdot\mathbf{a}''(u)=0$.
Furthermore, $\phi'''(u_s)\leqslant0$ because upon
taking the derivative of $\mathbf{a}'\cdot\mathbf{a}''(u)=0$, one gets 
$\mathbf{a}'\cdot\mathbf{a}'''(u)= - ||\mathbf{a}''||^2$ so that $\phi'''(u) = -  ||\mathbf{a}''||^2/2$. 
Regarding the Taylor expansion of $a'_i(u)$, the first term is $a'_i(u_s)=n_i$ which is suppressed by the 
$^{\scriptscriptstyle(TT)}$ projection operator
(it amounts to a gauge term \cite{DV}). To summarize, we therefore need to Taylor-expand the phase to the third order and $a'_i(u)$
to the first order: 
\be
\label{ }
\left\{
\begin{array}{ll}
\phi(u)\simeq\phi(u_s)+\frac{\phi'''(u_s)}{6}(u-u_s)^3 \\
\\
a'_i(u)\simeq a'_i(u_s) + (u-u_s) a''_i(u_s) \, ,
\end{array} \right.
\ee

\begin{figure}[h]
\begin{minipage}[t]{0.48\textwidth}
   \includegraphics[scale=0.4]{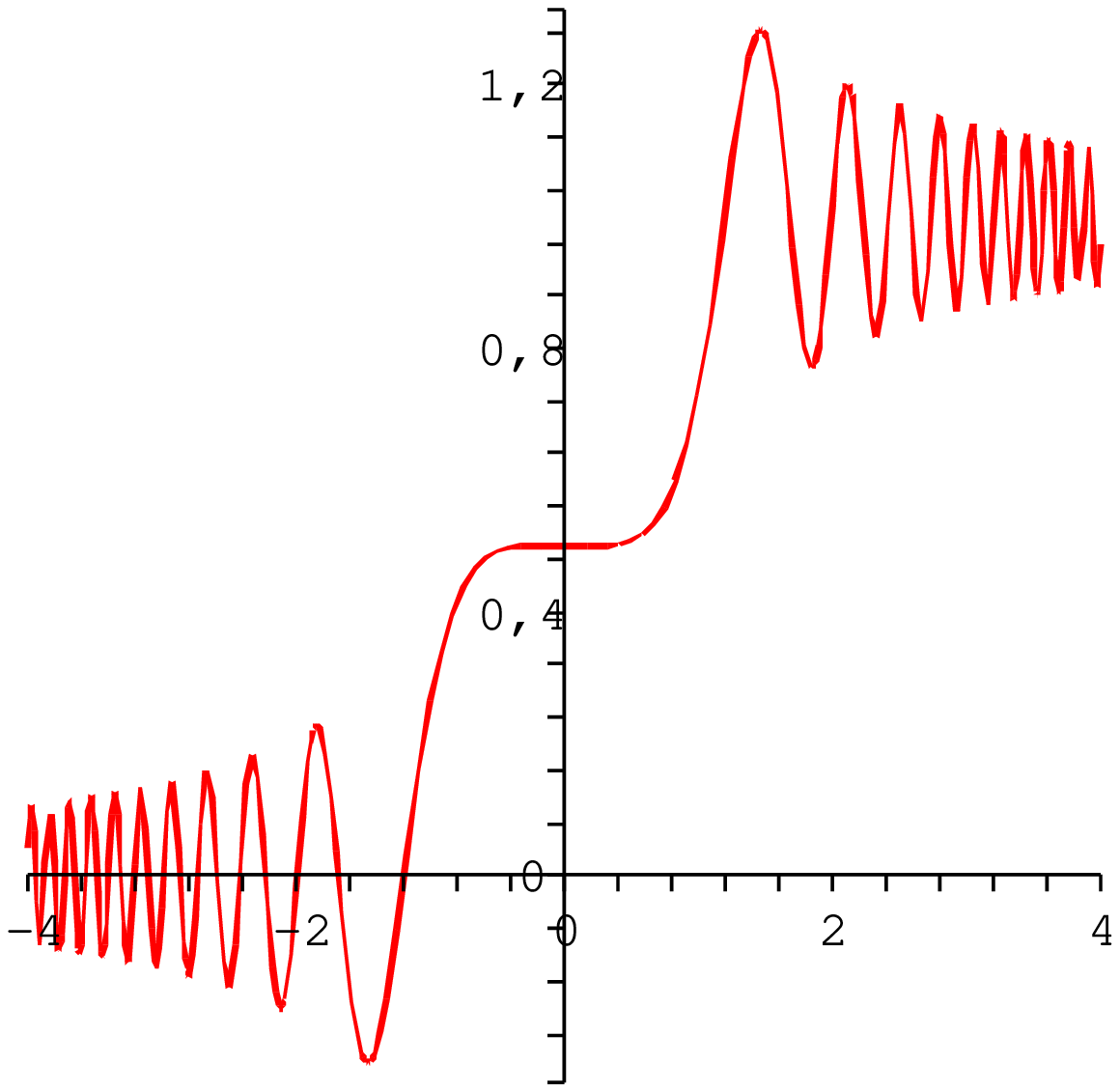} 
\end{minipage}
\begin{minipage}[t]{0.48\textwidth}
   \includegraphics[scale=0.4]{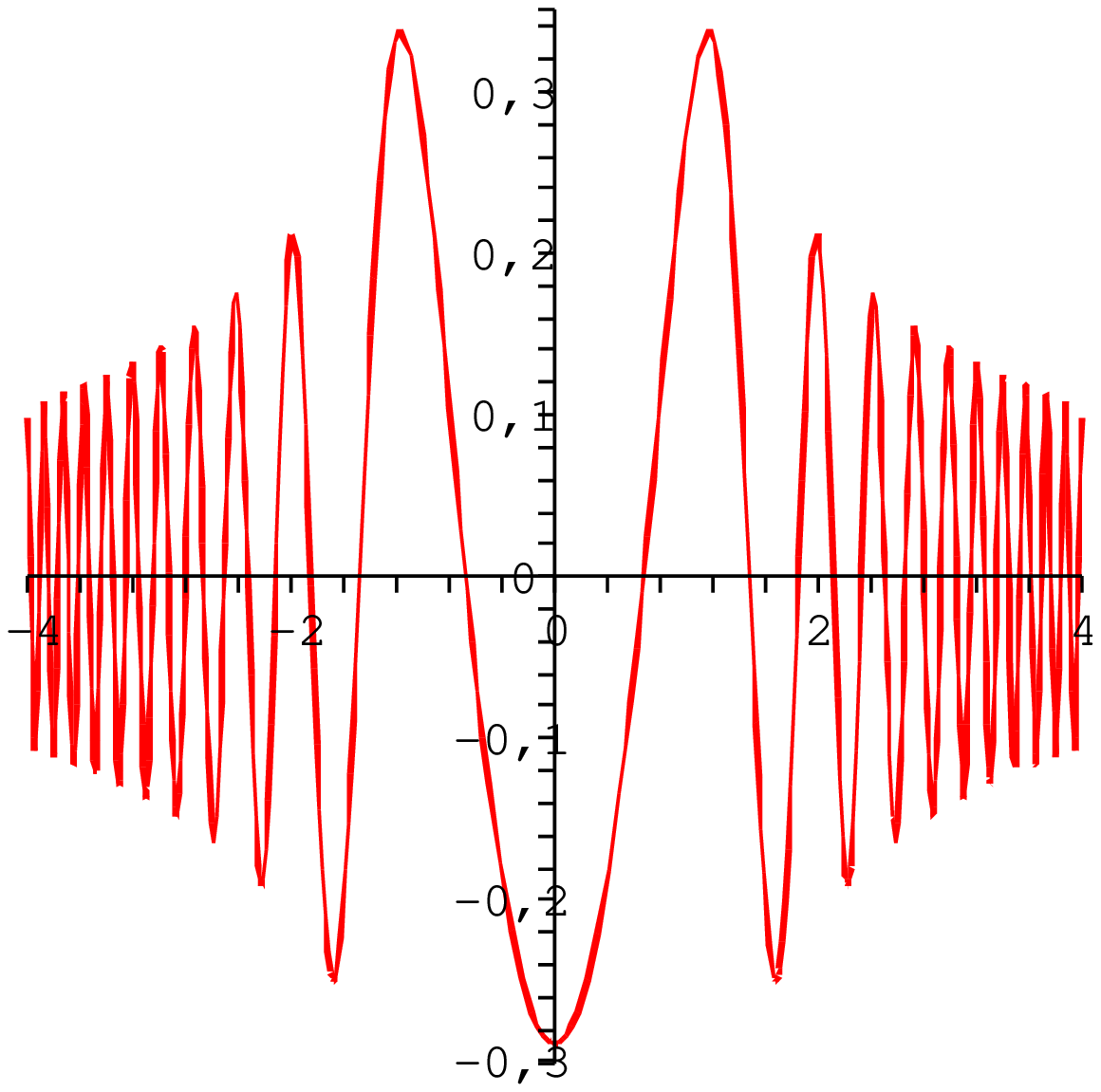} 
\end{minipage}
\caption{\bf Real and Imaginary parts of $B(w_A)$ as a function of $w_A$}
\label{B}
\end{figure}

Thus, up to a gauge term proportional to $n_i$,
\bea
\nonumber 
I_i & \simeq & e ^{-i\omega \phi(u_s)} \int _{u_A (v)} ^{u_B (v)} (u-u_s)
a''_i(u_s) e ^{-\frac{i\omega}{6}\phi'''(u_s)(u-u_s)^3}\d u
\\ 
\nonumber
 & \simeq & -e ^{-i\omega \phi(u_s)} a''_i(u_s) \Bigg( 
\frac{6}{\omega|\phi'''(u_s)|} \Bigg)^{2/3}\int _{w_B} ^{w_A} w e ^{-i w^3}
 \d w \qquad 
 %\qquad 
\left[w=-\left(\frac{\omega|\phi'''(u_s)|}{6}\right)^{1/3}(u-u_s)\right]
 \\
  \nonumber
 & \simeq & -e ^{-i\omega \phi(u_s)}a''_i(u_s) \Bigg( 
\frac{6}{\omega|\phi'''(u_s)|} \Bigg)^{2/3} \int _{-\infty} ^{\infty} 
w e ^{-i w^3}\d w 
 \\
 & \simeq &  \frac{1}{\omega ^{2/3}} a''_i(u_s) e ^{-i\omega \phi(u_s)}\Bigg( \frac{6}{|\phi'''(u_s)|} \Bigg)^{2/3}  \left( 
\frac{i}{\sqrt{3
}}\Gamma(\frac{2}{3})  \right)  \ ,
 \label{final}
\eea
where we have used $\int _{-\infty} ^{\infty} w e ^{-iw^3}dw=-\frac{i}{\sqrt{3}}\Gamma(\frac{2}{3})$.
We note that in going from the second to the third line, we have
assumed that the saddle point lies {\it far} from the boundaries:
\be
\label{ }
|u_s-u_J(v)||\omega\phi'''(u_s)| ^{1/3}\gg1 \ ,
\ee
thus allowing the domain of integration to be extended from $-\infty$ to 
$+\infty$. {\vf This will {\it not} always be the case, however, and the general result reads}
\be
%\begin{split}
%I_{saddle}(v)={\rm gauge~term} + \Bigg[\frac{1}{\omega ^{2/3}} &a''_i(u_s) 
%e ^{\frac{i \omega}{2}(u_s -  \mathbf{n}\cdot\mathbf{a}(u_s))} \Big( 
%\frac{12}{\mathbf{n}\cdot\mathbf{a}'''(u_s)} \Big) ^{2/3}\\
%&\times\frac{2}{3}i\sin(\pi/3)\Gamma(\frac{2}{3})C(w_A(v),w_B(v))\Bigg]\\
%\end{split}
I_{i}^{saddle}(v)={\rm gauge~term} + \Bigg[\frac{1}{\omega ^{2/3}} a''_i(u_s)
e ^{\frac{i \omega}{2}(u_s -  \mathbf{n}\cdot\mathbf{a}(u_s))} \left( 
\frac{12}{|\n \cdot \a'''|} \right) ^{2/3} \left( \frac{i}{\sqrt{3}} \Gamma(\frac{2}{3})\right) C(w_A,w_B)\Bigg]
\label{Isaddle}
\ee
with
\be 
C(w_A,w_B) = \frac{\int _{w_B} ^{w_A}w\  e ^{-i w^3}dw}{\int _{-\infty} 
^{\infty}w\  e ^{-i w^3}dw} = B(w_A) - B(w_B), \quad
B(w_J) = \frac{\int _{-\infty} ^{w_J}w\  e ^{-i w^3}dw}{\int _{-\infty} 
^{\infty}w\  e ^{-i w^3}dw}\ , \label{C} 
\ee
and
\be\nonumber
%w_A(v)  =  \left(\frac{\omega \phi'''(u_s)}{6}\right)^{1/3}(u_A(v)-u_s) &,&
%w_B(v) =  -\left(\frac{\omega \phi'''(u_s)}{6}\right)^{1/3}(u_s-u_B(v)) \ .
w_A(v)  =  \left(\frac{\omega |\n \cdot \a'''|}{12}\right)^{1/3}(u_s - u_A(v)), \quad
w_B(v) =  -\left(\frac{\omega |\n \cdot \a'''|}{6}\right)^{1/3}(u_B(v) - u_s).
\ee
%%%DS%%%
The behavior of the function $C(w_A,w_B)$ is studied in detail in Appendix A.
{\vf Its important features are the following,}
\begin{itemize}
\item when the saddle point {\vf lies well in between the upper and lower bounds of the $u$-integral}
 $C(w_A(v),w_B(v))$ obviously reduces to $1$, as in (\ref{final}).
\item when the saddle point is located far outside the domain of integration then 
$C(w_A(v),w_B(v))$ reduces to zero.
\item {\vf when the saddle point is near $u_A(v)$ (resp $u_B(v)$), 
$w_A(v)$ (resp $w_B(v)$) is of order one. 
$C(w_A(v),w_B(v))$ then reduces to $B(w_A(v))$ (resp $B(w_B(v))$) given in figure \ref{B}. }
We note that $B(w_A) \rightarrow 1$ for large values of $w_A$, with the envelope of its 
oscillations decreasing as $1/w$.
\end{itemize}

{\vf The entire dependence on $v$ in (\ref{Isaddle}) is contained in $C(v)=C(w_A(v),w_B(v))$, 
which is a smoothed version of the step function 
$\theta(u_s-u_A(v))\theta(u_B(v)-u_s)$ (cf Appendix A). 
This enters the integral over $v$ as follows.} Let $v_{A,s}$ (resp.~$v_{B,s}$) be 
the value of $v$ for which $u_A(v)=u_s$ (resp. $u_B(v)=u_s$). Since
$u_A$ and $u_B$ are decreasing functions of $v$, $C(w_A(v),w_B(v))$ is 
actually a smoothed version of the step function $\theta(v-v_{A,s})
\theta(v_{B,s}-v)$. {\vf But we ought to verify whether $C$ is rapidly }
varying (in the vicinity of $v_{A,s}$ and $v_{B,s}$) compared to the phase 
appearing in $e ^{-\frac{i \omega}{2}(v +  \mathbf{n}\cdot\mathbf{b}(v))}$. 
In {\vf typical
% (TH: instead of reasonable?
} cases where $|\frac{du_A}{dv}|\approx1$,$|\frac{du_B}{dv}|
\approx 1$, {\vf one has $\frac{dB}{dw}(w\approx w_A(v)) \sim {\cal O}(1)$} in the region of interest (see Figure \ref{B}). Hence

%%%DS%%%mods
\be
\label{dCdv1}
\frac{dC}{dv}(v\approx v_{A,s})\approx\underbrace{\frac{dB}{dw}(w\approx 
w_A(v))}_{\approx1}\Big(\omega|\phi'''(u_s)|\Big)^{1/3}\frac{du_A}{dv}
\approx\Big(\omega|\phi'''(u_s)|\Big)^{1/3}
\ee
For strings that are not too wiggly we expect $|\phi'''| \sim ||\a''||^2 \approx1/L^2$ so that
\be
\label{dCdv2}
\frac{dC}{dv}(v\approx v_{A,s})\approx\Big(\frac{\omega}{L^2}\Big)^{1/3}
\ll\omega
\ee
since we assumed $\omega\gg1/L$. Hence $C(v)$ enters as a multiplicative, slowly varying amplitude in the integral over 
$v$\footnote{Strictly speaking, we need to prove that $I(v)b_i'(v)$ is slowly varying 
in the sense that $\frac{d}{dv}(I(v)b_j'(v))\frac{1}{I(v)b_j'(v)} \ll \omega$. 
After applying the derivative one is left with two 
contributions: the first one is $\frac{b_j''(v)}{b_j'(v)}$ which, as before 
for the $a'$ term, is of order $\frac{1}{L} \ll \omega$. The second one is 
$\frac{I'}{I}=\frac{C'}{C}$ which, according to what we just said, is  $\ll 
\omega$.}. {\vf In particular it does not introduce a boundary term in the integral over $v$ (unless of 
course $du_B/dv$ diverges, see below), which implies junctions do not radiate spontaneously.}

\section{Full high frequency behaviour of $T_{ij}$}

{\vf We are now in a position to perform the remaining integral in (\ref{nonfstressE2}) to obtain the high frequency behavior of the various contributions to the stress energy source of GW bursts.}
%%%DS%%%corrected ref
The analysis of the previous section shows that,
in all cases, the integral over $v$ is {\vf of exactly the}  same type as the integral
$I_i(v)$ over $u$. We can therefore apply the same methods. 
{\vf Generalizing the definition we used above in the saddle point case we denote by $v_A(u)$ }
the value of $v$ for which $u_A(v)=u$ (and the dual definition for B). 
{\vf Since we are interested in the high frequency 
regime}, we list the different contributions with increasing powers of 
$1/\omega$. 

\subsection{Contributions in $1/\omega^{4/3}$} \label{4/3}

\begin{itemize}
\renewcommand{\labelitemi}{$\bullet$}

\item \underline{saddle point in the $u$ integral at $u_s$/ saddle point in 
the $v$ integral at $v_s$ (standard cusp)}
\vskip .3cm
The contribution from the integral over $u$ can be 
treated as a slowly varying amplitude. Hence the integral over $v$ is formally of the same type as the $u$-integral,
with the product $C(v)b'_j(v)$ now acting as the slowly varying amplitude. The derivative 
of this evaluated at $v_s$ contains a term proportional to $b'_j(v_s)$, which 
is a gauge term for the gravitational waves, and a term proportional to 
$C(v_s)b''_j(v_s)$. Our final result reads

% {\bf TH: where's the gauge term or is this the TT projection of the stress energy. In general where do we discuss this projection?}
\begin{equation}
\begin{split}
T _{ij}\approx \frac{\mu}{\omega ^{4/3}} & a''_i(u_s)b''_j(v_s) 
e ^{\frac{i \omega}{2}[u_s -  v_s - \mathbf{n}\cdot(\mathbf{a}(u_s)
+\mathbf{b}(v_s))]} 
\\ & \times \Big( 
\frac{12}{|\mathbf{n}\cdot\mathbf{a}'''(u_s)|} \Big) ^{2/3}  \Big( 
\frac{12}{\mathbf{n}\cdot\mathbf{b}'''(v_s)} \Big) ^{2/3}
%\\&\times
 \frac{1}{12}\Gamma^2(\frac{2}{3})C(w_A,w_B)\\
%&\times e ^{\frac{i \omega}{2}[u_s -  v_s - \mathbf{n}\cdot(\mathbf{a}(u_s)
%+\mathbf{b}(v_s))]} \ ,\\
\end{split}
\end{equation}
with
%%%DS%%%signs and modulus
\be\nonumber
\mathbf{n}  = \mathbf{a}'(u_s) = -\mathbf{b}'(v_s)
\ee
and
\be \nonumber
w_A  =  \left(\frac{\omega |\mathbf{n}\cdot\mathbf{a}'''(u_s)|}{12}\right)^{1/3}
(u_s- u_A(v_s)), \quad
w_B  = - \left(\frac{\omega |\mathbf{n}\cdot\mathbf{a}'''(u_s)|}{12}\right)^{1/3}
(u_B(v_s) - u_s).
\ee

This is the case of a cusp which emits in the 
direction given by $\mathbf{a}'(u_s)$. If the cusp occurs away from the junction, $C$ reduces to $1$ and we recover the standard
result of Damour and Vilenkin \cite{DV}.
\end{itemize}

\subsection{Contributions in $1/\omega^{5/3}$} \label{5/3}

This is the first case in which we find novel contributions specific to strings with junctions.

\begin{itemize}
\renewcommand{\labelitemi}{$\bullet$}

\item \underline{discontinuity in $a'_i$ at some $u_*$ / saddle point in the 
$v$ integral at $v_s$ (standard kink)}\\
\vskip .3cm
This is the standard case of a left-moving kink propagating on the string and 
emitting in the fan of directions $\mathbf{n}=-\mathbf{b}'(v)$ generated by 
the right moving waves:
\begin{equation}
\label{leftkink}
%%%DS%%%sin replaced, bracket fixed
\begin{split}
T _{ij}\approx \frac{\mu}{\omega ^{5/3}} & \Bigg(\frac{a^{\prime} 
_{i}(u_* ^{+})}{1 -  \mathbf{n}\cdot\mathbf{a}^{\prime}(u_*^{+})} -  
\frac{a^{\prime} _{i}(u_*^{-})}{1 -  
\mathbf{n}\cdot\mathbf{a}^{\prime}(u_*^{-})}\Bigg)b''_j(v_s) 
\Big( \frac{12}{\mathbf{n}\cdot\mathbf{b}'''(v_s)} \Big) ^{2/3}\\
&\times \frac{1}{2 \sqrt{3}} \Gamma(\frac{2}{3})\overline{C(w_A,w_B}  )e 
^{\frac{i \omega}{2}[u_* -  v_s - \mathbf{n}\cdot(\mathbf{a}(u_*)
+\mathbf{b}(v_s))]} \ ,\\
\end{split}
\end{equation}
with
%%%DS%%%removed, but added comment below
% \footnote{Notice the change of sign in the definition of $w_A$ and 
%$w_B$ compared to the study of the integral over $u$ due to the fact that 
%$\mathbf{n}\cdot\mathbf{b}'''(v_s)\geqslant0$. This choice was made so as to 
%recover a positive $w_A$ when $v_A(u_*)\leqslant v_s$ and induces a 
%conjugation of the complex factor $C$}
\be \nonumber
\mathbf{n} =-\mathbf{b}'(v_s)
\ee
and
\be \nonumber
w_A=\left(\frac{\omega\mathbf{n}\cdot\mathbf{b}'''(v_s)}{12}\right)
^{1/3}(v_s-v_A(u_*)), \quad
w_B=-\left(\frac{\omega \mathbf{n}\cdot
\mathbf{b}'''(v_s)}{12}\right)^{1/3}(v_B(u_*)-v_s).
\ee
%%%DS%%%added
In (\ref{leftkink}), $\overline{C}$ is the complex conjugate of $C$ (induced from the fact that in the Taylor expansion of the phase of the $v$ integral, we now have a {\it positive} $\phi'''(v)$, and hence this is equivalent to changing $i$ to $-i$ in all integrals.)

In order to illustrate how the presence of the function $C(w_A,w_B)$ 
modifies the case of a standard loop with no junction, let us consider a 
left moving kink propagating between B and A and ask 
in which directions $\mathbf{n}$ (described by points on the 
3D unit sphere) it emits. The set of points 
$\left\{ -\mathbf{b}'(v) \right\} $ draws a curve $\mathcal{C}$ on the Kibble-Turok sphere, as illustrated 
in figure \ref{KTsphere}. Away from $\mathcal{C}$, $T _{ij}(\mathbf{n}) 
\simeq 0$ whereas on the curve the amplitude is given by \eqref{leftkink}. 
The {\vf interval}
between positions $v_A(u_*)$ and $v_B(u_*)$ on the curve contains the right 
moving waves $-\mathbf{b}'(v)$ effectively seen by the kink while it 
propagates, so we expect to have an emission only in or around these 
directions. On this interval  - and not too close to the 
endpoints - we have $C\simeq1$. Far from this {\vf  interval\footnote{We note that although this is not taken into account by \eqref{leftkink}, $T _{ij}$ smoothly vanishes when 
$\mathbf{n}$ moves away from $\mathcal{C}$.}} (in blue on fig  \ref{KTsphere}), $C\simeq 0$. 
{\vf Finally, $C$ smoothly decreases from $1$ to $0$ in a short interval $\Delta v$ of a few $(L^2/\omega)^{1/3}$ near $v_A(u_*)$ and $v_B(u_*)$. Hence the overall probability that a GW burst comes with a value of $C$ significantly different from one 
is of the order $(L \omega)^{-1/3} \ll 1$. We conclude therefore that up to small corrections the predictions of \cite{DV} apply to GW bursts from kinks on cosmic strings with junctions.}

\begin{figure}[h]
\begin{center}
      \includegraphics[scale=0.45]{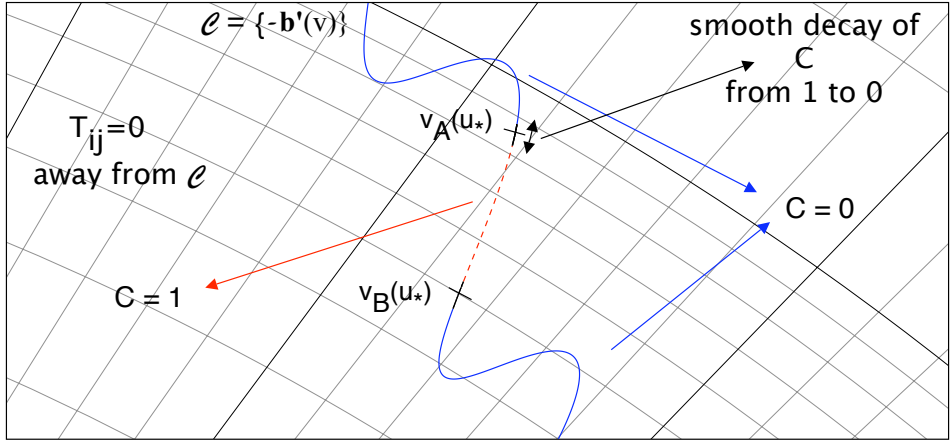}\\
\caption{\bf Directions of emission of a propagating kink on the Kibble-Turok 
sphere }
\label{KTsphere}
\end{center}
\end{figure}

%%%%

\item \underline{saddle point in the $u$ integral at $u_s$/ discontinuity of 
$\mathbf{b}'_j$ at some $v_*$ (standard kink)}\\
\vskip .3cm
This is a situation that {\vf mirrors} the previous one, namely a right-moving kink propagating on the string and 
emitting in a fan-like set of directions given by $\mathbf{n}=\mathbf{a}'$. 
The formulae are similar to those above.

%%%%

\item \underline{boundary term (A) in the integral over  $u$ / saddle point in the 
second integral at $v_s$}\\
\vskip .3cm
Inserting (\ref{Iedge}) into (\ref{nonfstressE2}), we obtain:
\begin{equation}
\label{ }
T _{ij}(\omega,\omega \mathbf{n}) = \frac{i\mu}{2\omega} 
\int _{-\infty} ^{\infty} \frac{b^{\prime} 
_{j}(v)a^{\prime} _{i}(u _{A}(v))}{1 -  \mathbf{n}\cdot\mathbf{a}^{\prime}(u 
_{A}(v))} e ^{-\frac{i \omega}{2}\Big[\big(v +  
\mathbf{n}\cdot\mathbf{b}(v)\big)-\big(u _{A}(v) -  
\mathbf{n}\cdot\mathbf{a}(u _{A}(v)\big)\Big] }\; \d v 
\end{equation}
where
 $\frac{b^{\prime} _{j}(v)a^{\prime} _{i}(u _{A}(v))}{1 -  
\mathbf{n}\cdot\mathbf{a}^{\prime}(u _{A}(v))}$ is a slowly varying amplitude 
compared to the {\vf rapidly} varying phase
\begin{equation}
\tilde{ \phi}(v)\equiv \frac{1}{2}\Big[\big(v +  \mathbf{n}\cdot\mathbf{b}(v)\big)
-\big(u_{A}(v) -  \mathbf{n}\cdot\mathbf{a}(u _{A}(v)\big)\Big] \ .
\end{equation}
The saddle point condition is that
 \begin{equation}
\tilde{ \phi}'(v_s)=\frac{1}{2}\Big[ \underbrace{1+\mathbf{n}\cdot\mathbf{b}
^{\prime}(v_s)}_{\geqslant 0} +\underbrace{\underbrace{\frac{du_A}{dv}(v_s)}
_{\leqslant 0}\underbrace{(\mathbf{n}\cdot\mathbf{a}^{\prime}(u_B(v_s))-1)}
_{\leqslant 0}}_{\geqslant 0}\Big]=0\, .
 \end{equation}

Since both terms in the sum are positive, they both need to vanish at 
$v_s$. Moreover $\mathbf{n}\cdot\mathbf{a}^{\prime}(u_A(v_s))\not=-1$,
otherwise we would have a cusp at the junction. Hence {\vf we are led to the following conditions}:
%%%DS%%%added implication into formula
\begin{equation}
\label{condjuncsad}
\left\{
\begin{array}{rll}
\mathbf{n}\cdot\mathbf{b}^{\prime}(v_s) & =&-1 \qquad \Rightarrow \qquad \n = -\mathbf{b}'(v_s) \\
\frac{du_A}{dv}(v_s) & = & 0
\end{array} \right.
\end{equation}

%$\mathbf{n}\cdot\mathbf{b}^{\prime}(v_s)=-1$ and $\frac{du_A}{dv}(v_s)=0$. 

The second condition {\vf means} the string is
expanding at the speed of light at junction A. The other case where the string 
expands at the speed of light at junction B would correspond to 
$\frac{du_B}{dv}=-\infty$, and thus to a discontinuity in the $v$ integral.
We will treat this separately below.

The amplitude of this contribution can be calculated using the saddle 
point treatment described previously (namely a Taylor expansion of the amplitude 
and the phase). We have
\begin{equation}
\tilde{ \phi}''(v_s)=\frac{1}{2}\Bigg[\underbrace{\mathbf{n}\cdot\mathbf{b}''(v_s)}_{= 
0}+\underbrace{\frac{d^2 u_A}{dv^2}(v_s)
\big(\mathbf{n}\cdot\mathbf{a}'(u _{A}(v_s))-1 \big)}_{= 0}
+\underbrace{\mathbf{n}\cdot\mathbf{a}''(u _{A}(v))
\Big(\frac{du_A}{dv}(v_s)\Big)^2}_{= 0}\Bigg] \ .
\end{equation}
The first term vanishes because of the gauge conditions on the 
worldsheet, and the last term vanishes due to the second condition in 
\eqref{condjuncsad}. The second term also vanishes. Indeed, since $\frac{d u_A}{dv}(v) 
\leqslant 0$ and $\frac{d u_A}{dv}(v_s) = 0$, $\frac{d u_A}{dv}(v)$ 
reaches its maximum at $v_s$. 
Hence we need to Taylor expand the phase to third order:
\begin{equation}
\tilde{ \phi}'''(v_s) = \frac{1}{2}\Bigg[\mathbf{n}\cdot\mathbf{b}'''(v_s)+ 
\frac{d^3 u_A}{dv^3}(v_s)\big(\mathbf{n}\cdot\mathbf{a}'(u _{A}(v_s))-1 \big) 
\Bigg]
\end{equation}
On the other hand, Taylor expansion of the amplitude to first order gives
\begin{equation}
b^{\prime} _{j}(v)\frac{a^{\prime} _{i}(u _{A}(v))}{1 -  
\mathbf{n}\cdot\mathbf{a}^{\prime}(u _{A}(v))} \approx 
\hbox{gauge~term} + (v-v_s) b'' _{j}(v_s)\frac{a^{\prime}_{i}(u_{A}(v_s))}{1 
-  \mathbf{n}\cdot\mathbf{a}^{\prime}(u _{A}(v_s))} \ ,
\end{equation} 
where the gauge term is proportional to $b^{\prime} _{j}(v_s) = - n_j$.
%\begin{equation}
%\begin{split}
%b^{\prime} _{j}(v)\frac{a^{\prime} _{i}(u _{A}(v))}{1 -  
%\mathbf{n}\cdot\mathbf{a}^{\prime}(u _{A}(v))} \approx & 
%\underbrace{b^{\prime} _{j}(v_s)\frac{a^{\prime} _{i}(u _{A}(v_s))}{1 -  
%\mathbf{n}\cdot\mathbf{a}^{\prime}(u _{A}(v_s))}}_{gauge~term} \\
%&+(v-v_s)\Big(b'' _{j}(v_s)\frac{a^{\prime} _{i}(u _{A}(v_s))}{1 -  
%\mathbf{n}\cdot\mathbf{a}^{\prime}(u _{A}(v_s))}+  \underbrace{b^{\prime} 
%_{j}(v_s)\frac{d}{dv}\big(\frac{a^{\prime} _{i}(u _{A}(v))}{1 -  
%\mathbf{n}\cdot\mathbf{a}^{\prime}(u _{A}(v))}
%\big)(v=v_s)}_{gauge~term~and~=0}  \Big)
%\end{split}
%\end{equation}

Hence the leading term of the energy-momentum tensor is given by
%%%DS%%%sin replaced
\begin{equation}
\label{ }
\begin{split}
T _{ij}(\omega,\omega \mathbf{n}) \approx \frac{\mu}{\omega ^{5/3}}  & b'' 
_{j}(v_s)\frac{a^{\prime} _{i}(u _{A}(v_s))}{1 -  
\mathbf{n}\cdot\mathbf{a}^{\prime}(u _{A}(v_s))}\Big(\tilde{ \phi}'''(v_s)\Big) 
^{2/3}\\
& \times \frac{6 ^{2/3}}{2 \sqrt{3}} \Gamma(\frac{2}{3}) e ^{-\frac{i 
\omega}{2}\Big[\big(v_s +  \mathbf{n}\cdot\mathbf{b}(v_s)\big)
-\big(u _{A}(v_s) -  \mathbf{n}\cdot\mathbf{a}(u _{A}(v_s)\big)\Big] }
\end{split}
\end{equation}

This contribution corresponds to the {\vf situation in which} the string expands at the speed 
of light at junction A and emits in a direction 
corresponding to $\mathbf{n}=-\mathbf{b}^{\prime}(v_s)$, with an 
amplitude proportional to $1/\omega ^{5/3}$.
%\vskip .3cm
We now turn to the case where the string expands at the speed of light 
at junction B. \\

\item \underline{saddle point in the $u$ integral at $u_s$/ string expanding 
at the speed of light at junction B}\\
\vskip .3cm
In this case $du_B/dv$ diverges, which gives rise 
to a discontinuity-like contribution in the integral over $v$.
From (\ref{nonfstressE2}) and (\ref{Isaddle}) we have
%%%DS%%%sin replaced, mod added
\begin{equation}
\label{ }
\begin{split}
T _{ij}(\omega,\omega \mathbf{n}) \approx \frac{\mu}{4}  & \frac{1}{\omega 
^{2/3}} a''_i(u_s) e ^{\frac{i \omega}{2}(u_s -  
\mathbf{n}\cdot\mathbf{a}(u_s))} \Big( 
\frac{12}{|\mathbf{n}\cdot\mathbf{a}'''(u_s)|} \Big) 
^{2/3}\frac{i}{\sqrt{3}}\Gamma(\frac{2}{3})\\
&\times \int _{-\infty} ^{\infty} C(v) b^{\prime} _{j}(v) e ^{-\frac{i 
\omega}{2}(v +  \mathbf{n}\cdot\mathbf{b}(v))} \d v \ ,
\end{split}
\end{equation}
%%%DS%%%I would still like to do more things from here, but no time. 
where $C(v)=C(w_A(v),w_B(v))$ was defined in (\ref{C}).
We are interested in cases in which a discontinuity in $C$ (or at least a 
variation faster than that of the phase in the exponential, $|dC/dv|\gg\omega$) is induced by the 
divergence of $du_B/dv$ or equivalently $dw_B/dv$
at some $v_*$.  Yet, to achieve such a discontinuity in 
$C$, we need $v_*$ to satisfy
\begin{equation}
\label{wBordre1}
|w_B(v_*)|\lesssim1 \ .
\end{equation}
Indeed, in this region we have $\partial C/\partial w_B\approx1$ 
whereas elsewhere $C$ does not vary with $w_B$ at leading order so that 
a {\vf strong} variation of $w_B(v)$ would have no effect on $C(v)$. Physically, this 
requirement means that the saddle point condition is satisfied near the 
junction at a moment when $\dot{s}_B$ reaches $1$.

From now on we assume that  \eqref{wBordre1} is satisfied.  For $|dC/dv|\gg\omega$, we can treat $C$ as being discontinuous at $v_*$ where $dC/dv$ diverges.
%
%Even though $dC/dv$ {\vf is} diverging at $v_*${\vf this} does not mean $C$ is 
%discontinuous at $v_*$, it can be treated as such if $|dC/dv|\gg\omega$. {\vf TH: I don't understand this sentence. DS: tried to clarify}
In this case, the leading contribution is the same as that of a 
discontinuous $C$ with a gap $\Delta C(v*)$ equal to the variation of the 
true function $C$ over the interval where $|dC/dv|\gg\omega$:

\begin{equation}
\label{avantcalculdeltaC}
\begin{split}
T _{ij}(\omega,\omega \mathbf{n}) \approx \frac{\mu}{\omega ^{5/3}}  & a''
_i(u_s) \frac{b'(v_*)}{1+\mathbf{n}\cdot\mathbf{b}'(v_*)}\Big( 
\frac{12}{|\mathbf{n}\cdot\mathbf{a}'''(u_s)|} \Big) ^{2/3} \times \Delta C(v_*)\\
&\times -\frac{1}{2\sqrt{3}}\Gamma(\frac{2}{3}) e 
^{\frac{i \omega}{2}\big(u_s -v_* -  \mathbf{n}\cdot(\mathbf{a}(u_s)
+\mathbf{b}(v_*))\big)}
\end{split}
\end{equation}

We therefore need to evaluate the order of magnitude of $\Delta C(v_*)$. Let 
us start by determining the interval around $v_*$ over which 
$|dC/dv|\gg\omega$. As was already explained in (\ref{dCdv1}) and (\ref{dCdv2}), 
$|dC/dv|\approx |du_B/dv|(\omega/L^2)^{1/3}$ so the condition 
$|dC/dv|\gg\omega$ translates into $|du_B/dv|\gg(\omega L)
^{2/3}$. Since $|du_B/dv|=(1+\dot{s}_B)/(1-\dot{s}_B)\approx 
2/(1-\dot{s}_B)$ (because $\dot{s}_B\approx1$ in the region of interest),
this finally yields the condition
\begin{equation}
\label{condclosespeedlightB}
1-\dot{s}_B\ll \frac{1}{(\omega L) ^{2/3}} \ .
\end{equation}
We therefore need to determine how $\dot{s}_B(t)$ behaves around a point $t_*$ 
where $\dot{s}_B(t_*)=1$. {\vf To this end we} investigate the dynamics
of a junction using the equations recalled in Section \ref{CSwJ} 
(we momentarily drop the B index in equation (\ref{s1}), and assume the above discussion applies to string $1$, namely $\dot{s}_1(t_*) =1$, while strings $2$ and $3$ are assumed {\it not} to be expanding at the speed of light.)
%, ($\dot{s}_{2,3}(t_*) \ll 1$)}).  
For convenience define $v _{2*}=s_2(t_*)-t_*$ and $v _{3*}=s_3(t_*)-t_*$. 
%
% Suppose that at some $t_*$, we  have $\dot{s}_1(t_*)=1$. 
%
Then, from (\ref{s1}),
$c_1(t_*)=1$ so that $\b_2'(v_{2*})=\b_3'(v_{3*})\equiv \B$.
% are equal, with say $\mathbf{b}'_2(v_2)=\mathbf{b}'_3(s_3(t_*)-t_*)=\B$.  
Expansion of $\mathbf{b}'_2(v_2)$ and $\mathbf{b}'_3(v_3)$ around $v _{2*}$ and 
$v _{3*}$ respectively then yields
\begin{eqnarray}
\nonumber \mathbf{b}'_2(v_2) & \approx & \B + \mathbf{b}''_2(v 
_{2*})( v_{2}-v _{2*}) + \mathbf{b}'''_2(v _{2*})\frac{( v_{2}-v _{2*})^2}{2}\\
\nonumber \mathbf{b}'_3(v_3) & \approx & \B+ \mathbf{b}''_3(v 
_{3*})( v_{3}-v _{3*}) + \mathbf{b}'''_3(v _{3*})\frac{( v_{3}-v _{3*})^2}{2} \, ,
\end{eqnarray}
%{\bf DS: corrected a 3 to 2 above, and sometimes changed $s_2(t)-t$ to $v_2(t)$, and same for $3$}
where we have used the fact that $\mathbf{b}''_2(v _{2*})\cdot\B=\mathbf{b}''_3(v _{3*})\cdot\B=0$ because of the gauge constraints 
on the worldsheets.  Thus

\begin{equation}
\label{c1dev}
\begin{split}
%c_1(t)=\mathbf{b}'_2(v_2) \cdot \mathbf{b}'_3(v_3)\approx 1 + & \frac{(v_3(t)-v_{3*})^2}{2} \mathbf{b}'''
%_3(v _{3*})\cdot\B+ \frac{(v_2(t)-v_{2*})^2}{2}\mathbf{b}'''
%_2(v _{2*})\cdot\B\\
%& +(v_2(t)-v_{2*})(v_3(t)-v_{3*})\mathbf{b}''_2(v 
%_{2*})\cdot\mathbf{b}''_3(v _{3*})
c_1(t)\approx 1 + & \frac{(s_3(t)-t-v_{3*})^2}{2} \mathbf{b}'''
_3(v _{3*})\cdot\mathbf{B}+ \frac{(s_2(t)-t-v_{2*})^2}{2}\mathbf{b}'''
_2(v _{2*})\cdot\mathbf{B}\\
& +(s_2(t)-t-v_{2*})(s_3(t)-t-v_{3*})\mathbf{b}''_2(v 
_{2*})\cdot\mathbf{b}''_3(v _{3*})
\end{split}
\end{equation}

We can also expand 
$s_2(t)-t\approx v_{2*}+(\dot{s}_2(t_*)-1)(t-t_*)$ and 
$s_3(t)-t\approx v_{3*}+(\dot{s}_3(t_*)-1)(t-t_*)$ 
%$v_2(t)\approx v_{2*}+(\dot{s}_2(t_*)-1)(t-t*)$ and 
%$v_3(t)\approx v_{3*}+(\dot{s}_3(t_*)-1)(t-t*)$ 
where generically 
$(\dot{s}_2(t_*)-1)$ and $(\dot{s}_3(t_*)-1)$ are of order $1$. Then, using 
the fact that the three scalar products involved in \eqref{c1dev} are of order 
$1/L^2$, we can write, around $t_*$
\begin{equation}
c_1(t)-1\approx -\frac{1}{L^2}(t-t_*)^2
\end{equation}
where the minus sign ensures that $c_1(t)\leqslant1$. Plugging this into 
equation (\ref{s1}), we finally get the expansion around $t_*$
\begin{equation}
1-\dot{s}_1(t)\approx \frac{1}{L^2}(t-t_*)^2 \ .
\end{equation}

Therefore, \eqref{condclosespeedlightB} {\vf holds} over an interval of time 
around $t_*$ of size $\Delta t$ of order $\Delta t \approx L/(\omega L) 
^{1/3}=L^{2/3}\omega ^{-1/3}$. Since $u_B=s_B(t)+t$, the variation of 
$u_B$ during this interval $\Delta t$ is of the same order as $\Delta t$ i.e. 
$\Delta u_B \approx L^{2/3}\omega ^{-1/3}$ and so the variation of 
$w_B$ is $\Delta w_B \approx (\omega L^{-2})^{1/3} \Delta u_B 
\approx  (\omega L^{-2})^{1/3} L^{2/3}\omega ^{-1/3}
\approx 1$. Finally, since we assumed that $w_B(v_*)$ is in the region where 
$\partial C/ \partial w_B \approx1$, this yields
\begin{equation}
\Delta C(v_*) \approx 1
\end{equation}\\
To conclude, if $\mathbf{n}=\mathbf{a}^{\prime}(u_s)$ and  
$du_B/dv$ diverges at some $v_*$ such that \newline
$|w_B(v_*)|=\Big( 
\frac{\omega|\mathbf{n}\cdot\mathbf{a}'''(u_s)|}{12} \Big) ^{1/3}|u_B(v_*)-u_s|\lesssim1$ then the integral over $v$ receives a discontinuity-like 
contribution leading to
\begin{equation}
\label{ }
\begin{split}
T _{ij}(\omega,\omega \mathbf{n}) \approx \frac{\mu}{\omega ^{5/3}}  & 
a''_i(u_s) \frac{b'_j(v_*)}{1+\mathbf{n}\cdot\mathbf{b}'(v_*)}\Big( 
\frac{12}{|\mathbf{n}\cdot\mathbf{a}'''(u_s)|} \Big) ^{2/3} \\
&\times -\frac{1}{2\sqrt{3}}\Gamma(\frac{2}{3}) \times 
\Delta C(v_*) e ^{\frac{i \omega}{2}\big(u_s -v_* -  \mathbf{n}
\cdot(\mathbf{a}(u_s)+\mathbf{b}(v_*))\big)}
\end{split}
\end{equation}
where with respect to \eqref{avantcalculdeltaC} we now {\vf write} $\Delta C(v_*)$ on 
the second line to indicate that this is a numerical factor of order $1$. The 
string expands at the speed of light at junction B.\\

%with
%\begin{equation}
%\label{}
%\left\{
%\begin{array}{rll}
%\mathbf{n}\cdot\mathbf{a}^{\prime}(u_s) & =&1 \\
%\frac{du_s}{dv}(v_*) & = & -\infty \qquad i.e. \qquad \dot{s}_B(t_*)=1
%\end{array} \right.
%\end{equation}

\end{itemize}

\subsection{Contributions in $1/\omega^2$} \label{2}

\begin{itemize}
\renewcommand{\labelitemi}{$\bullet$}

\item \underline{Discontinuity in $a'_i$ at some $u_*$ / Discontinuity in 
$b'_j$ at some $v_*\in[v_A(u_*),v_B(u_*)]$}

\begin{equation}
\begin{split}
T _{ij}\approx \frac{\mu}{\omega^2}  & 
\Bigg(\frac{a^{\prime} _{i}(u_* ^{+})}{1 -  
\mathbf{n}\cdot\mathbf{a}^{\prime}(u_*^{+})} -  \frac{a^{\prime} 
_{i}(u_*^{-})}{1 -  \mathbf{n}\cdot\mathbf{a}^{\prime}(u_*^{-})}\Bigg)\\
&  \times\Bigg(\frac{b^{\prime} _{j}(v_* ^{+})}{1 +  
\mathbf{n}\cdot\mathbf{b}^{\prime}(v_*^{+})} -  \frac{b^{\prime} 
_{j}(v_*^{-})}{1 +  \mathbf{n}\cdot\mathbf{b}^{\prime}(v_*^{-})}\Bigg)    
e ^{\frac{i \omega}{2}[u_* -  v_* - \mathbf{n}\cdot(\mathbf{a}(u_*)
+\mathbf{b}(v_*))]}
\end{split}
\end{equation}

This is the situation where a left moving and a right moving kink meet during 
their propagation on the string and emit in all directions.\\

%%%%%

\item \underline{Discontinuity in $a'_i$ at some $u_*$ /  boundary term in the 
integral over $v$}

%As we already mentioned the inside integral has a dependence on $v$ of the 
%form
%$\mathbf{1}_{u_A(v)<u_*<u_B(v)}=\mathbf{1}_{v_A(u_*)<v<v_B(u_*)} $ which 
%induces  two edge terms in the $v$ integral. We give the one corresponding to 
%the $B$ bound. 
\begin{equation}
\begin{split}
T _{ij}\approx -\frac{\mu}{\omega^2}  & 
\Bigg(\frac{a^{\prime} _{i}(u_* ^{+})}{1 -  
\mathbf{n}\cdot\mathbf{a}^{\prime}(u_*^{+})} -  
\frac{a^{\prime} _{i}(u_*^{-})}{1 -  
\mathbf{n}\cdot\mathbf{a}^{\prime}(u_*^{-})}\Bigg)\\
& \times \Bigg(\frac{b^{\prime} _{j}(v_B(u_*))}{1 +  
\mathbf{n}\cdot\mathbf{b}^{\prime}(v_B(u_*))} \Bigg)
e ^{\frac{i \omega}{2}[u_* -  v_B(u_*) - 
\mathbf{n}\cdot(\mathbf{a}(u_*)+\mathbf{b}(v_B(u_*))]}
\end{split}
\end{equation}
This corresponds to a left-moving kink passing through a junction and 
emitting in all directions in space.\\

%%%%%%

\item \underline{boundary term (A or B) in the integral over  $u$ /discontinuity 
of $b_j'$ at some $v_*$}\\
\begin{equation}
\begin{split}
T _{ij}\approx -\frac{\mu}{\omega^2}  & \Bigg(\frac{b^{\prime} _{j}(v_* 
^{+})}{1 +  \mathbf{n}\cdot\mathbf{b}^{\prime}(v_*^{+})} -  \frac{b^{\prime} 
_{j}(v_*^{-})}{1 +  \mathbf{n}\cdot\mathbf{b}^{\prime}(v_*^{-})}\Bigg)\\
& \times \Bigg(\frac{a^{\prime} _{i}(u_A(v_*))}{1 -  
\mathbf{n}\cdot\mathbf{a}^{\prime}(u_A(v_*))} \Bigg)
e ^{\frac{i \omega}{2}[u_A(v_*) -  v_* - \mathbf{n}\cdot(\mathbf{a}(u_A(v_*))
+\mathbf{b}(v_*)]}
\end{split}
\end{equation}
This describes a right moving kink passing through a junction (A here) and emitting in all directions in space. As one expects, the symmetry with respect to the case of a left moving kink passing through B is 
recovered despite the explicit breaking of symmetry made by choosing to 
integrate on $u$ first.\\

%%%%%

\end{itemize}

\section{Conclusion}

The transverse traceless projections of the various contributions to the stress-energy tensor found in Section V fully determine the GW emission from cusps and kinks in the local wave zone of the source through eq \eqref{Tijtohij}. The observed signal is obtained by parallel propagation of the gravity waves in a cosmological background, along the null geodesic followed by the GW. This gives rise to the usual redshifting of time intervals between emission and reception, which in the Fourier domain corresponds to $f_{em} = (1+z) f_{rec}$. One obtains the following order-of-magnitude estimates for the logarithmic Fourier transform of the observed amplitude of individual GW bursts emanating from the various high frequency sources discussed here \cite{DV,DV1,Xavier2},
\be \label{obssignal}
h(f_{rec}) \sim \frac{G \mu L}{((1+z) Lf_{rec})^{\alpha}}\frac{1+z}{t_0 z}
\ee
where $t_0$ denotes the present age of the universe. Most importantly, the exponent $\alpha$ in this expression is determined by the high frequency behavior $\sim \omega^{-(1+\alpha)}$ of the stress-energy sources calculated above. 

In the time domain \eqref{obssignal} corresponds to a signal of the form $\sim |t-t_c|^{\alpha}$, where $t_c$ is the arrival time of the center of the burst. The total GW signal of a superstring loop consists of the sum of the various bursts and a slowly varying component due to the low frequency modes of the strings. Despite its vanishing at $t=t_c$, bursts are distinguishable from the slowly varying component because the curvature associated with \eqref{obssignal} diverges as $ \sim |t-t_c|^{\alpha-2}$, exhibiting clearly the spiky nature of GW bursts\footnote{In practice, the observer will never lie exactly in the direction of emission. However the signal is detectable inside a small cone around this direction. This introduces a high frequency cut-off on the waveform or, equivalently, a smoothing of the signal around $t_c$ in the time domain.}.

We are now in a position to summarize the modifications of the GW signal from bursts induced by the presence of junctions,
with respect to the case of standard loops. As expected, {\it away from the junctions} we recover the same GW signal as Damour and Vilenkin \cite{DV,DV1}, both for a kink propagating on a string - for which $\alpha=2/3$ and the emission occurs in 
a fan-like set of directions - as well as for a cusp, where $\alpha=1/3$ and the emission occurs in a small cone around a particular
direction. 

However for bursts from sources near junctions we find the predicted GW amplitude is modified by a smooth correction factor $C$. This means that for GW bursts from cusps and kinks near junctions one effectively measures a combination of the tension $\mu$ and $C$. Since these are rather improbable events, however, the overall effect of this on the (statistical) predictions of bursts from cosmic superstrings is likely to be small. The smoothness of the correction factor also implies that despite their spiky nature, junctions do not radiate spontaneously (at least for strings without structure, as considered here). 

Furthermore, the presence of junctions gives rise to several novel sources of GW bursts, most notably from
\begin{itemize}
\item a string expanding at the speed of light at the level of a junction, which emits in a specific direction 
with an amplitude \eqref{obssignal} with $\alpha=2/3$.
\item a kink passing through a junction, which emits in all directions with an amplitude \eqref{obssignal} with $\alpha=1$.
\end{itemize}
Finally, the situation where a left-moving and a right-moving kink pass through each other during their propagation remains 
unchanged and has $\alpha=1$.

To translate our results into observable waveforms one ought to sum the individual contributions from cusps and strings in a cosmological network of string loops. However the rate of occurrence of events of different types strongly depends on the dynamics of the loops and on the evolution of the network, which is currently poorly understood. 
%Suppose one wishes to detect at least one event per year. Then one needs to observe up to a certain redshift. 
This gives rise to significant uncertainties in the resulting GW signal. Indeed, an increase in the number of events of a given type lowers the threshold redshift for observation and hence reduces the expected dilution of the signal. The individual contribution with the smallest value of $\alpha$ therefore need not necessarily provide the dominant contribution to the observed signal. 

In particular, concerning the case at hand, one might expect junctions to enhance the number density of kinks whereas it appears unlikely that the rate of occurrence of cusps is significantly affected by the presence of junctions \cite{BBHS2}. This would mean that even though individual GW bursts from cusps are stronger, the increased number density of kinks on loops with junctions might compensate for the difference in strength. Since we have identified GW bursts from events involving kinks that are specific to strings with junctions, this would provide a promising route to observationally distinguish between gauge theory cosmic strings with no junctions and superstrings.

%We see that, in order to obtain detailed predictions for observable waveforms, one needs to combine our results with a precise undertanding of the string network and of its dynamics. This is far from being achieved in the case of strings with junctions. One thus has to resort to simplifying assumptions. We defer such a task to future work.

%As for standard loops, the strongest emission comes from cusps. However, if 
%one wishes to detect such an emission, one needs to take into account the 
%propagation of the gravitational wave from the loop to us which results in a 
%dilution of the signal. In order to ensure a fixed minimal rate of detection, 
%one needs to observe up to a certain redshift (he further we observe, the 
%more events you see) and therefore has to deal with a weaker signal. The 
%other crucial point then is the number of events per unit of time. [see DV]. 

%\begin{acknowledgments}
\section*{Acknowledgements}
This work has been supported by the LISAScience grant ANR-07-BLAN-0339 of the Agence Nationale
de la Recherche.  D.A.S.~would like to thank the University of Geneva, Switzerland, for hospitality whilst this work was in progress. T.H. thanks the Perimeter Institute (Waterloo, Canada) for hospitality during the completion of this work.  
%\end{acknowledgments}

\appendix

\section{Detailed study of the function $C(v) \equiv  C(w_A,w_B)$}

In \eqref{Isaddle}, we gave the general result for the
integral \eqref{Iv} due to a saddle point:
\be
I_{i}^{saddle}(v)={\rm gauge~term} + \Bigg[\frac{a''_i(u_s)}{\omega ^{2/3}}
e ^{\frac{i \omega}{2}(u_s -  \mathbf{n}\cdot\mathbf{a}(u_s))} \left( 
\frac{12}{|\n \cdot \a'''|} \right) ^{2/3} \left( \frac{i}{\sqrt{3}} \Gamma(\frac{2}{3})\right) C(w_A,w_B)\Bigg]
\ee
where
\be
C(w_A,w_B) = \frac{\int _{w_B} ^{w_A}w\  e ^{-i w^3}dw}{\int _{-\infty} 
^{\infty}w\  e ^{-i w^3}dw} = B(w_A) - B(w_B), \quad 
B(w_J) = \frac{\int _{-\infty} ^{w_J}w\  e ^{-i w^3}dw}{\int _{-\infty} 
^{\infty}w\  e ^{-i w^3}dw}
\ee
and
\be
w_A(v)  =  \left(\frac{\omega |\n \cdot \a'''|}{12}\right)^{1/3}(u_s - u_A(v)), \quad 
w_B(v) =  -\left(\frac{\omega |\n \cdot \a'''|}{6}\right)^{1/3}(u_B(v) - u_s).
\ee
%%\be
%I_{saddle}(v)={\rm gauge~term} + \frac{1}{\omega ^{2/3}} a''_i(u_s) 
%e ^{\frac{i \omega}{2}(u_s -  \mathbf{n}\cdot\mathbf{a}(u_s))} \Big( 
%\frac{12}{\mathbf{n}\cdot\mathbf{a}'''(u_s)} \Big) ^{2/3}\times\frac{2}{3}i
%\sin(\pi/3)\Gamma(\frac{2}{3})\times C(w_A(v),w_B(v))\\
%\ee
%with the following definitions
%\bea
%C(w_A,w_B) = B(w_A) - B(w_B) & , & B(w_J) = \frac{\int _{-\infty} ^{w_J}w\  
%e ^{-i w^3}dw}{\int _{-\infty} ^{\infty}w\  e ^{-i w^3}dw} \ , \\
%w_A(v)  =  (\frac{\omega\phi'''(u_s)}{6})^{1/3}(u_A(v)-u_s) &,&
%w_B(v) =  -(\frac{\omega\phi'''(u_s)}{6})^{1/3}(u_s-u_B(v)) \ .
%\eea
The function $C$ depends on $v$ because the position of the saddle point relative to the bounds $u_A(v)$ and $u_B(v)$ of the 
integral does. This Appendix contains a detailed analysis of $C(v)$, whose behavior needs to be understood in order to determine the saddle point contributions in the three cases listed under \eqref{C}.\\ \\
%
%
%\subsection{Decrease as $\frac{1}{w_A}$ of the envelope of $B(w_A)=
%\frac{\int _{-\infty} ^{w_A}w\  e ^{-i w^3}dw}{\int _{-\infty} ^{\infty}w\  
%e ^{-iw^3}dw}$ }
%
%
\indent {\bf Envelope of oscillations of $B$}

When the saddle point is far from $w_B$, we can write
$C(w_A,w_B) \sim B(w_A)$,
where $B(w_A)$ is displayed in Figure \ref{B}. We first show that the envelope 
of the oscillations of $B$ scales as $1/w_A$.

Consider for example the real part
$\Re(B(w_A))=\alpha\int _{-\infty} ^{w_A} w 
\sin(w^3)\d w$, where \newline
$\alpha=-i/\big(\int _{-\infty} ^{\infty} w e ^{-iw ^3} \d w 
\big) $, which oscillates around its asymptotic value of 1 with decreasing amplitude.
Let $w_k=(\pi k) ^{1/3}$ be the values where its derivative $(\alpha w 
\sin(w^3))$ vanishes. These points correspond to the relative maxima (for odd 
values of $k$) and minima (for even values of $k$) of $\Re(B(w_A))$. Then, 
because $w _{k+1}-w _k $ is small, the amplitude of the oscillations around a 
certain $w_k$ (say $k$ even for example) is given (for large values of $k$) 
by 
\be
\frac{B(w _{k+1}) - B(w_k)}{2}=\frac{\alpha}{2}\int _{w_k} ^{w_{k+1}} w 
\sin(w^3)dw  \sim \frac{\alpha}{2} \Big[\frac{-\cos(w^3)}{3w}\Big]_{w_k} ^{w_{k+1}} 
\sim \frac{\alpha}{3 w_k}.
\ee

We infer that $B(w)$ is of order $1/w$ in the limit $w \rightarrow -
\infty$. In particular, if $w_B$ is negative and large in absolute value, 
we have $C(w_A,w_B)\approx B(w_A)$. 
\\ 
\\
\indent {\bf Contributions to \eqref{Isaddle} due to a saddle point inside 
$\left[u_A(v),u_B(v)\right]$}

%\subsection{Contributions to \eqref{Isaddle} due to a saddle point inside 
%$\left[u_A(v),u_B(v)\right]$}
Following the discussion in section III, we investigate the contribution due 
to a saddle point lying between $u_A(v)$ and $u_B(v)$. This corresponds to 
studying $C(v)$ between the points $v_{A,s}$ and $v_{B,s}$ defined at the end 
of Section IV. Since $\phi'''(u_s)\leqslant0$, $u_s\in\left[u_A(v),u_B(v)
\right]$ translates into $w_A(v)\geqslant0$ and $w_B(v)\leqslant0$. We need to 
distinguish between two regimes:

{\em a. Saddle point far from both bounds: $|u_s-u_J(v)||\omega\phi'''(u_s)| 
^{1/3}\gg1$}

Then we have $w_A(v)\gg1$ and $|w_B(v)|\gg1$ so
 according to the discussion above, $C(v)\approx1$ up to corrections of the 
order of $1/w_A,1/w_B\ll1$

{\em b. Saddle point near one of the boundaries (e.g. A):  $|u_s-u_A(v)||\omega
\phi'''(u_s)| ^{1/3}\lesssim1$}

First, note that the saddle point cannot be close to both bounds at the same 
time:  $(\omega\phi''')^{1/3}$ is of the order of $\omega L ^{-2}\gg1/L$ so 
$|u_s-u_A(v)|\ll L$; since $|u_A(v)-u_B(v)|$ is of the order of $L$,  
$|u_s-u_B(v)|$ has to be of the order of $L$ too so we still have $|w_B(v)|
\gg1$. 

Therefore, up to a negligible correction of the order of $1/w_B$, $C(w_A(v),
w_B(v))\approx B(w_A(v))$. This time $w_A$ is of order $1$ (and still 
positive). The behaviour of $B(w_A)$ is plotted in Figure \ref{B}. The 
contribution in this case is of the same order as the one far from the bounds 
but the result is multiplied by the complex factor $B(w_A)$ of order $1$.
\\
\\
\indent {\bf Contributions to \eqref{Isaddle} due to a saddle point 
\emph{outside} $\left[u_A(v),u_B(v)\right]$}

Up to now we calculated the leading contribution from a saddle point present 
anywhere \emph{inside} the interval $[u_A(v),u_B(v)]$. Does this mean that for 
the values of $v$ for which the saddle point lies outside this interval, 
$I(v)$ discontinuously drops to zero?

We argued in Section \ref{sec:cond} that, if $I_i(v)$ obeys all conditions 
{\em (i)-(iii)}, then it is negligible because it vanishes exponentially in 
the infinite $\omega$ limit. However, since $\omega$ is actually large but 
finite, there might also exist non negligible contributions if $I(v)$ is only 
\emph{close to} satisfying these conditions. In particular, if some saddle 
point lies just outside the interval of integration, the derivative at the 
nearest bound is very close to vanishing so we can expect to have some 
contribution. Another way to say this is that the contribution comes from an 
interval around the saddle point of size $\approx (L^2/\omega) ^{1/3}$ which 
can intersect $\left[u_A(v),u_B(v)\right]$ even if the saddle point is outside 
it. Obviously, in the case where the saddle point is very far from the 
interval of integration, the contribution needs to be negligible.

{\em a. Saddle point outside $\left[u_A(v),u_B(v)\right]$ and near one 
of the boundaries (e.g. A): \newline $|u_s-u_A(v)||\omega\phi'''(u_s)| ^{1/3}\lesssim1$}.
Here, $w_B\leqslant0$. For the same reasons as before, we have $|w_B|\gg1$ and 
$C(w_A(v),w_B(v))\approx B(w_A(v))$. The only difference is that now 
$w_A\leqslant0$. This is why in Figure \ref{B} we also plotted $B$ for 
negative values of its argument. 

{\em b. Saddle point far from $\left[u_A(v),u_B(v)\right]$}

In this case, both $w_A$ and $w_B$ have the same sign and their absolute 
values are $\gg1$. According to the discussion in the first section of this 
appendix, we have at leading order $C(v)\approx0$ up to corrections in 
$1/w_A,1/w_B\ll1$.
\\
\\
\indent {\bf Summary}
%\subsection{Synthesis of the diferent cases: complete behaviour of the 
%function $C(v)$}

\begin{figure}[h]
%\hspace{-2.8cm}
\begin{minipage}[t]{0.40\textwidth}
   \includegraphics[scale=0.2]{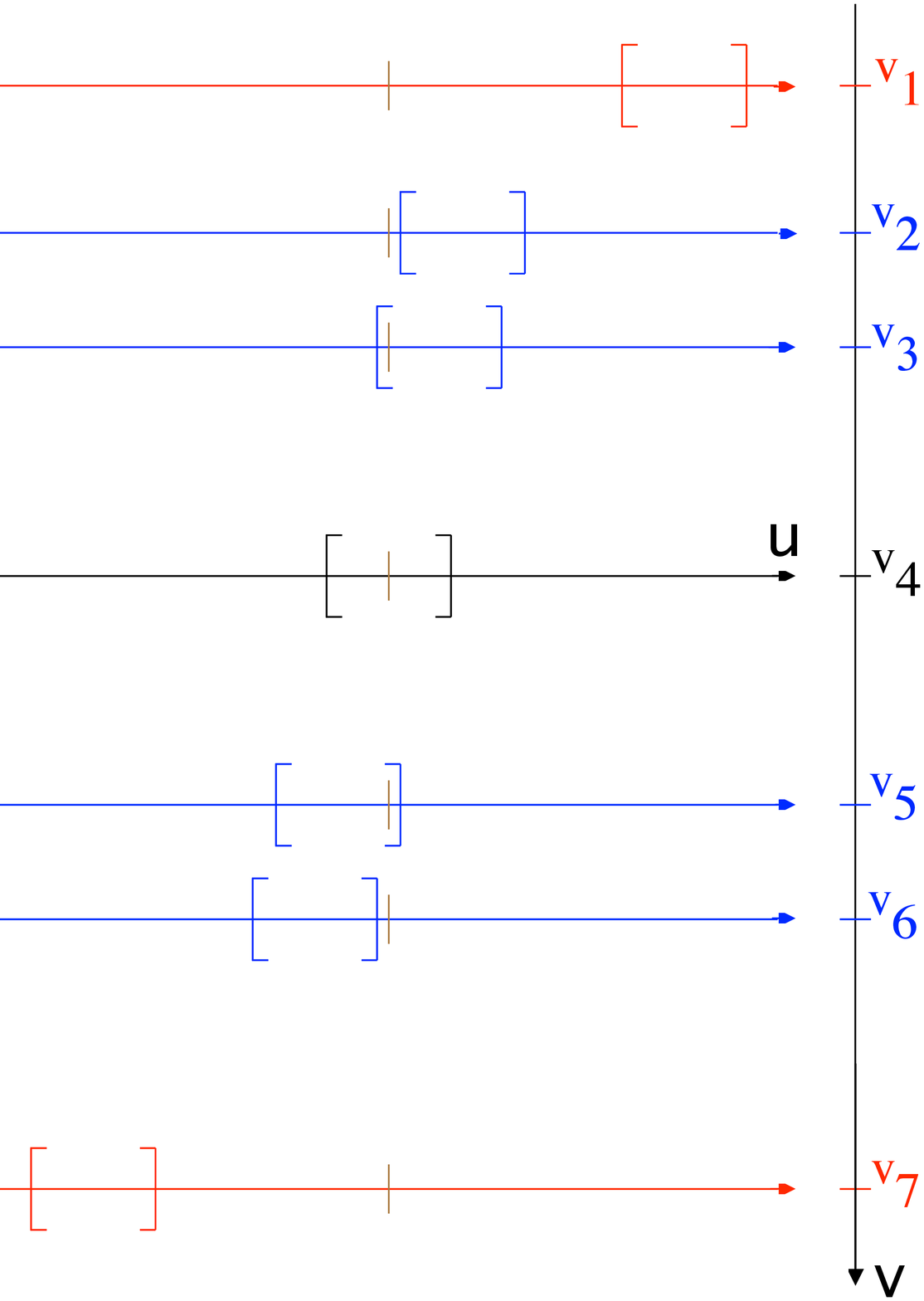}
\end{minipage}
\hspace{-2cm}
\begin{minipage}[t]{0.56\textwidth}
%\begin{sideways}
   \includegraphics[scale=0.28]{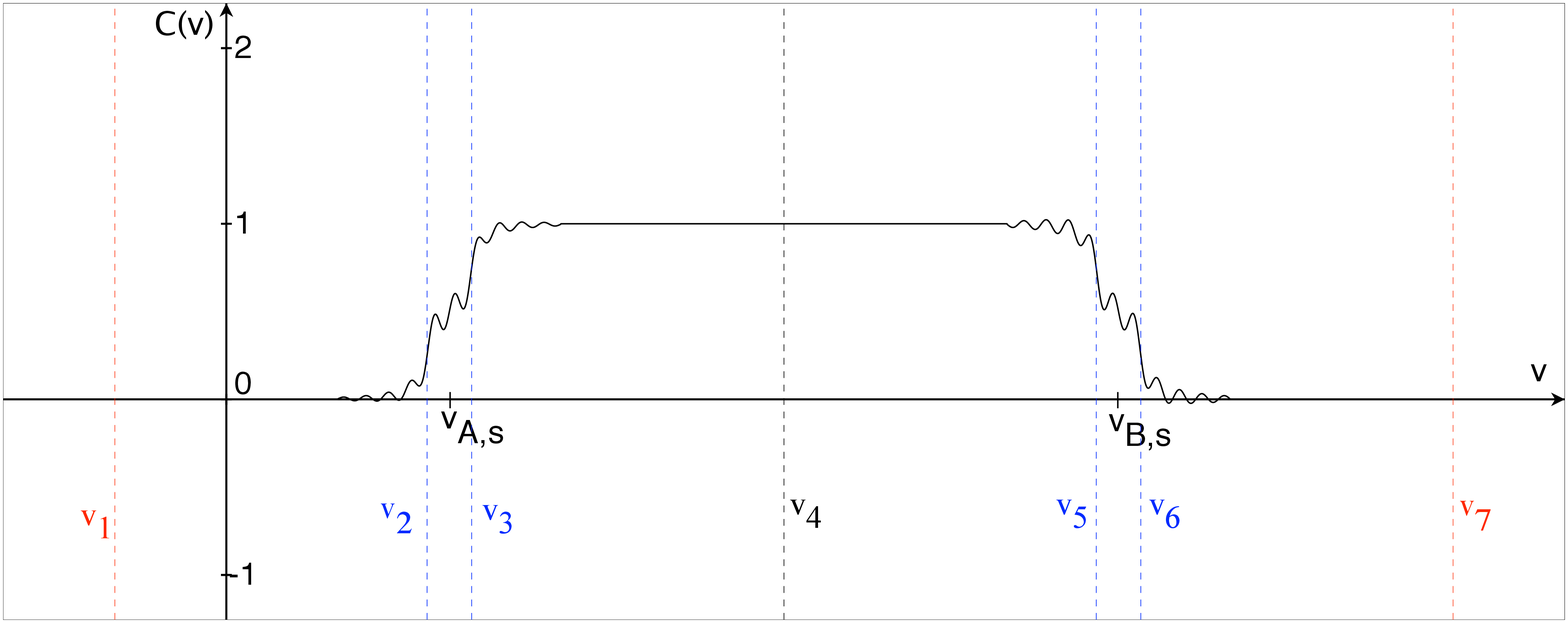}
% \end{sideways}
\end{minipage}
\caption{Typical behaviour of the function $C(v)$. The different cases are 
represented on the left panel. The left and right brackets respectively 
indicate the position of $u_A(v)$ and $u_B(v)$ on the $u$ axis while the brown 
tick is the position of $u_s$}
\label{figC}
\end{figure}

It should now be clear that $C(v)$ is a smoothed version of $\theta(v-v_{A,s})
\theta(v_{B,s}-v)$. In Figure \ref{figC} we sum up the different cases and 
show the typical behaviour of the function $C(v)$ provided that the derivatives
$\frac{du_J}{dv}$ are of order $1$ which is the generic case.
%\begin{figure}[h]
%\begin{center}
%   \includegraphics[scale=0.2]{differentcaseswithouttext.eps}
%\end{center}
%\end{figure}
%\begin{figure}[h]
%\begin{center}
%%\begin{sideways}
%   \includegraphics[scale=0.28]{step.eps}
%% \end{sideways}
%\end{center}
%\caption{Typical behaviour of the function $C(v)$. The different cases are 
%represented on the left panel. The left and right brackets respectively 
%indicate the position of $u_A(v)$ and $u_B(v)$ on the $u$ axis while the brown 
%tick is the position of $u_s$}
%\label{figC}
%\end{figure}

\end{document}